\documentstyle[12pt,epsfig]{article}
 1
 1
 1

\begin{document}
\bibliographystyle{unsrt}
%
\def\lta{\;\raisebox{-.5ex}{\rlap{$\sim$}} \raisebox{.5ex}{$<$}\;}
\def\gta{\;\raisebox{-.5ex}{\rlap{$\sim$}} \raisebox{.5ex}{$>$}\;}
\def\grle{\;\raisebox{-.5ex}{\rlap{$<$}}    \raisebox{.5ex}{$>$}\;}
\def\legr{\;\raisebox{-.5ex}{\rlap{$>$}}    \raisebox{.5ex}{$<$}\;}

%
\def\r#1{\ignorespaces $^{\rm #1}$} 
\def\l#1{\ignorespaces $_{\rm #1}$} 

\newcommand{\ra}{\rightarrow}
\newcommand{\permille}{$^0 \!\!\!\: / \! _{00}$}
\newcommand{\dd}{{\rm d}}
\newcommand{\oal}{{\cal O}(\alpha)}%
\newcommand{\su}{$ SU(2) \times U(1)\,$}
 
\newcommand{\np}{Nucl.\,Phys.\,}
\newcommand{\pl}{Phys.\,Lett.\,}
\newcommand{\pr}{Phys.\,Rev.\,}
\newcommand{\prl}{Phys.\,Rev.\,Lett.\,}
\newcommand{\prep}{Phys.\,Rep.\,}
\newcommand{\zp}{Z.\,Phys.\,}
\newcommand{\sovjnp}{{\em Sov.\ J.\ Nucl.\ Phys.\ }}
\newcommand{\nuclinst}{{\em Nucl.\ Instrum.\ Meth.\ }}
\newcommand{\annp}{{\em Ann.\ Phys.\ }}
\newcommand{\intjmp}{{\em Int.\ J.\ of Mod.\  Phys.\ }}
 
\newcommand{\eps}{\epsilon}
\newcommand{\mw}{M_{W}}
\newcommand{\mww}{M_{W}^{2}}
\newcommand{\mbb}{m_{b \bar b}}
\newcommand{\mcc}{m_{c \bar c}}
\newcommand{\mh}{m_{H}}
\newcommand{\mhh}{m_{H}^2}
\newcommand{\mz}{M_{Z}}
\newcommand{\mzz}{M_{Z}^{2}}

\newcommand{\lra}{\leftrightarrow}
\newcommand{\tr}{{\rm Tr}}
 
\newcommand{\ie}{{\em i.e.}}
\newcommand{\cm}{{{\cal M}}}
\newcommand{\cl}{{{\cal L}}}
\def\Ww{{\mbox{\boldmath $W$}}}  
\def\B{{\mbox{\boldmath $B$}}}         
\def\nn{\noindent}

\newcommand{\sinsq}{\sin^2\theta}
\newcommand{\cossq}{\cos^2\theta}
\newcommand{\be}{\begin{equation}}
\newcommand{\beq}{\begin{equation}}
\newcommand{\eeq}{\end{equation}}
\newcommand{\ee}{\end{equation}}
\newcommand{\ba}{\begin{eqnarray}}
\newcommand{\ea}{\end{eqnarray}}
\newcommand{\beqn}{\begin{eqnarray}}
\newcommand{\eeqn}{\end{eqnarray}}
\newcommand{\bea}{\begin{eqnarray}}
\newcommand{\ena}{\end{eqnarray}} 
\newcommand{\eea}{\end{eqnarray}}

\newcommand{\nl}{\nonumber \\}
\newcommand{\eqn}[1]{Eq.(\ref{#1})}
\newcommand{\ibidem}{{\it ibidem\/},}
\newcommand{\into}{\;\;\to\;\;}
\newcommand{\wws}[2]{\langle #1 #2\rangle^{\star}}
\newcommand{\smod}{\tilde{\sigma}}
\newcommand{\dilog}[1]{\mbox{Li}_2\left(#1\right)}
\newcommand{\umu}{^{\mu}}
\newcommand{\cjg}{^{\star}}
\newcommand{\lgn}[1]{\log\left(#1\right)}
\newcommand{\si}{\sigma}
\newcommand{\sit}{\sigma_{tot}}
\newcommand{\sqs}{\sqrt{s}}
\newcommand{\sih}{\hat{\sigma}}
\newcommand{\sith}{\hat{\sigma}_{tot}}
\newcommand{\p}[1]{{\scriptstyle{\,(#1)}}}
\newcommand{\res}[3]{$#1 \pm #2~~\,10^{-#3}$}
\newcommand{\rrs}[2]{\multicolumn{1}{l|}{$~~~.#1~~10^{#2}$}}
\newcommand{\err}[1]{\multicolumn{1}{l|}{$~~~.#1$}}
\newcommand{\ru}[1]{\raisebox{-.2ex}{#1}}
\newcommand{\epem}{$e^{+} e^{-}\;$}
\newcommand{\epemt}{e^{+} e^{-}\;}
\newcommand{\eeah}{$e^{+} e^{-} \ra H \gamma \;$}
\newcommand{\eahnw}{$e\gamma \ra H \nu_e W$}

\newcommand{\thebb}{\theta_{b-beam}}
\newcommand{\pte}{p^e_T}
\newcommand{\ptH}{p^H_T}
\newcommand{\gag}{$\gamma \gamma$ }
\newcommand{\gam}{\gamma \gamma }

\newcommand{\aatoh}{$\gamma \gamma \ra H \;$}
\newcommand{\egam}{$e \gamma \;$}
\newcommand{\eat}{e \gamma \;}
\newcommand{\eaeh}{$e \gamma \ra e H\;$}
\newcommand{\eaehb}{$e \gamma \ra e H \ra e (b \bar b)\;$}
\newcommand{\egebb}{$e \gamma (g) \ra e b \bar b\;$}
\newcommand{\egecc}{$e \gamma (g) \ra e c \bar c\;$}
\newcommand{\eaebb}{$e \gamma \ra e b \bar b\;$}
\newcommand{\eaecc}{$e \gamma \ra e c \bar c\;$}
\newcommand{\aah}{$\gamma \gamma H\;$}
\newcommand{\zah}{$Z \gamma H\;$}
\newcommand{\pe}{P_e}
\newcommand{\pg}{P_{\gamma}}
\newcommand{\delbb}{\Delta m_{b \bar b}}

\begin{titlepage}
\rightline{ROME1-1165/97}
\rightline{NDU-HEP-97-EG01}
\rightline{February 1997}
\rightline{Revised July 1998}
\vskip 22pt 

\noindent
\begin{center}
{\Large \bf {\boldmath \zah} vertex effects in Higgs production \\
at future {\boldmath \egam} linear colliders. 
  }
\end{center}
\bigskip

\begin{center}
{\large 
E.~Gabrielli~$^{a},$
$\;\,$V.A.~Ilyin~$^b \;\,$
and $\;\,$B.~Mele~$^c$. 
} \\
\end{center}

\medskip\noindent
$^a$ University of Notre Dame, IN, USA \\
\noindent
$^b$ Institute of Nuclear Physics, Moscow State University, Russia \\
\noindent
$^c$ INFN, Sezione di Roma 1 and Rome University ``La Sapienza", Italy
\bigskip
\begin{center}
{\bf Abstract} \\
\end{center}
{\small 
One-loop production of a Higgs boson in \egam collisions at future 
accelerators is studied via the process \eaeh, for intermediate Higgs masses.  
Exact cross sections, including the possibility of longitudinally
 polarized initial beams, are presented.
Confirming previous estimates made in the Weizs\"acker-Williams
approximation, they are found to be
more than two orders of magnitude larger than the cross sections 
for the crossed process \eeah, in the  energy range $\sqs=(0.5\div 2)$ TeV.
We show that, not only \eaeh  has a similar
potential as the \aatoh process for testing the 
one-loop \aah  vertex, but,
by requiring a final electron tagged at large angle in \eaeh, 
the $He$ production provides an excellent 
way of testing the \zah vertex, too. 
Kinematical distributions for the \eaehb process with a tagged 
final electron are analyzed, and
strategies for controlling the main irreducible background are found.
Initial-state-radiation effects are checked to be within a few percent. 
}

\vskip 22pt 
\vfil
\noindent
e-mail: \\
egabriel@wave.phys.nd.edu, $\;$
ilyin@theory.npi.msu.su, $\;$
mele@roma1.infn.it \\ 

\end{titlepage}

\noindent
\section{Introduction}
\vskip 10pt
The Higgs boson sector is a crucial part of the Standard Model still
escaping direct experimental verification. 
Presently, we know that $\mh\gta 65$GeV \cite{higl}.
Once the Higgs boson will be discovered either at LEP2 or at LHC,  
testing the Higgs boson properties will be a central issue at future linear 
colliders.
In particular, an \epem collider with centre-of-mass (c.m.) energy $\sqs\simeq
(300\div 2000)$GeV and integrated luminosity ${\cal O}(100)$ fb$^{-1}$
will allow an accurate determination of the mass, couplings and 
parity properties of the Higgs particle \cite{saar,zerw}. Two further options
are presently considered for a high-energy linear collider, where
one or both the initial $e^+/e^-$ beams are replaced by  
photon beams induced by Compton backscattering of laser light on
the high-energy electron beams \cite{spec}. Then, one can study 
high-energy electron-photon and photon-photon collisions, 
where the initial photons are real, to a
good degree monochromatic, and have energy and luminosity
comparable to the ones of the parent electron beam \cite{mono}. 

In this paper, we analyse the Higgs production in \egam collisions 
through the process \eaeh.
This channel will turn out to be an excellent mean to test 
both the \aah and \zah one-loop couplings with high statistics.
Possible ways to test the couplings $ggH$, \aah and \zah have been 
extensively studied in the literature. These one-loop vertices,
because of the nondecoupling properties of the Higgs boson, are sensitive
to the contribution of new particles circulating in the loops,
even in the limit $M_{new}\gg \mh$ \cite{higg}.

While the $ggH$ vertex \cite{hggu} can be tested by the 
Higgs production via gluon-gluon
fusion at LHC, a measurement of the \aah and \zah  couplings
should be possible by the determination of the BR's for the decays
$H\ra \gamma \gamma$ \cite{haau,haad} and $H\ra \gamma Z$
\cite{hzad,haad} (see also \cite{holl}), respectively. The latter
statement holds
only for an intermediate-mass Higgs boson (i.e., for 90GeV$\lta \mh
\lta 140$ GeV), where both BR($H\ra \gamma \gamma$) and
BR($H\ra \gamma Z$) reach their maximum values, which is ${\cal O}(10^{-3})$.

Another promising way of measuring the \aah coupling for an
intermediate-mass Higgs boson will be realized through
Higgs production in \gag collisions \cite{aahu,aahd}. To this end,
the capability of tuning the \gag c.m. energy on the Higgs mass,
through  a good degree of the photons  monochromaticity, will be crucial
for not diluting too much  the $\gam \to H$ 
resonant cross section over the c.m. energy spectrum.

The process \eaeh, that we consider here, offers a further
interesting way of testing both the \aah and \zah
Higgs vertices. Indeed, we will show that, while the $\gamma$-exchange
\aah contribution is dominant in the total
cross section,  by requiring a large transverse momentum of the final
electron (or Higgs boson), one enhances the $Z$-exchange \zah contribution,
while keeping the corresponding rate still to an observable level. 
The further contribution given by the box diagrams with $W$ and $Z$ exchange 
survives  at large angles too, but is relatively less important.
Furthermore, while the  \aah and \zah channels increase logarithmically with
the c.m. collision energy, the contribution from 
boxes starts decreasing at $\sqs\gta 400$ GeV.

A further advantage of the \eaeh process with respect to the resonant
$\gam\to H$ production is that the former is much less crucially 
dependent on the tuning of the c.m. collision energy to $\mh$.
As a consequence, although  the resonant cross section 
$\si_{res}(\gam\to H)$ is in general
much larger than $\si($\eaeh), the effect of  the $\gamma$-spectrum smearing
can make the two rates of the same order of magnitude
(see also \cite{aahd}).

The cross section for the process \eaeh has previously been studied in the 
Weizs\"acker-Williams (WW) approximation \cite{wewi}, where the only channel
contributing is the (almost real) $\gamma$-exchange in the $t$-channel, 
induced by the \aah vertex \cite{ebol} (see also \cite{dicu}, where the 
pseudoscalar Higgs-boson production is considered). 
Although, as we will see, this method provides a rather good 
estimate of the \eaeh total cross sections, it is unable 
to assess the importance of the \zah (and box) effects. 
This we will address particularly in our exact treatment of \eaeh.

Although, cross sections for the process \eaeh are quite large also for heavy
Higgs bosons (e.g., $\sigma(\mh\simeq 400GeV)>1$ fb for $\sqs\gta 500$ GeV),
we will concentrate on the intermediate 
Higgs mass case. Hence, we will carry out a detailed analysis of 
the main background, assuming that the decay $H\to b \bar b$ is dominant.
 
In principle, the same physics could be tested in 
the crossed process, \eeah, which has been widely studied 
\cite{barr,abba,djou}.
Unfortunately, the \eeah channel suffers from small rates, which 
are further depleted at large energies by the $1/s$ behavior of the dominant
s-channel diagrams. Also, in this case, it is more difficult to separate
the \zah contribution on the basis of kinematical distributions.
As a consequence, if a $e\gamma$ option of the linear collider will be
realized with similar luminosity of the \epem option,
the \eaeh channel will turn out to be much more interesting than the 
process \eeah,
for finding possible deviations from  the standard-model 
one-loop Higgs vertices.

The plan of the paper is the following. In Section 2,
we present the analytical results for the complete helicity 
amplitudes of the
\eaeh process. In Section 3, numerical results for the exact total cross
section are given and compared to the ones corresponding to the 
tree-level Higgs production in \eahnw. Also, a discussion of the relative
importance of the different one-loop vertices and boxes in \eaeh is presented.
In Section 4, the rates for the main  background processes  
are estimated, and strategies for their control are suggested.
Initial-beam polarization effects are discussed in Section 5,
while , in Section 6, we estimate the influence of the Initial State Radiation
(ISR) on the above picture.
In Section 7, we discuss the expected precision on a measurement
of the \zah effects through \eaeh, and point out a possible strategy,
based on the angular asymmetry of the final electron, for
further optimizing the ratio $S/B$.
Finally, in Section 8, we draw our conclusions.
In the Appendix, we discuss some technical details of the computation.


\newpage
\noindent
\section{Helicity amplitudes}
\vskip 10pt
In this section we give the analytical expression for the matrix element
of the process 
\beq
e^{-}(k_1) \gamma(k_2) 
\rightarrow e^{-}(k_3) H(k_4)
\label{proc}
\eeq 
as a function of the initial electron and photon helicities,
where $k_i$ are the particles momenta. 
We calculate the amplitude in the {\it 't-Hooft-Feynman gauge} 
and in the chiral limit approximation for the electron mass.

\begin{figure}
\begin{center}
\begin{picture}(160,100)
\put(-30,-70){\epsfxsize=20cm \leavevmode \epsfbox{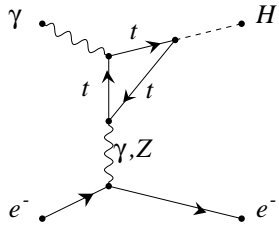} }
\end{picture} 
\end{center}
\caption{Feynman diagram with fermion triangle loop.}
\label{dgrm1}  
\end{figure}

\begin{figure}
\begin{picture}(160,100)
\put(0,-70){\epsfxsize=20cm \leavevmode \epsfbox{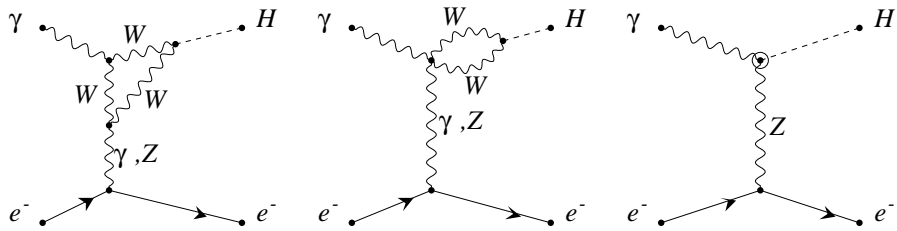} }
\end{picture} 
\caption{Feynman diagrams with $W$-triangle loop.}
\label{dgrm2}  
\end{figure}

\begin{figure}
\begin{picture}(160,100)
\put(-40,-70){\epsfxsize=20cm \leavevmode \epsfbox{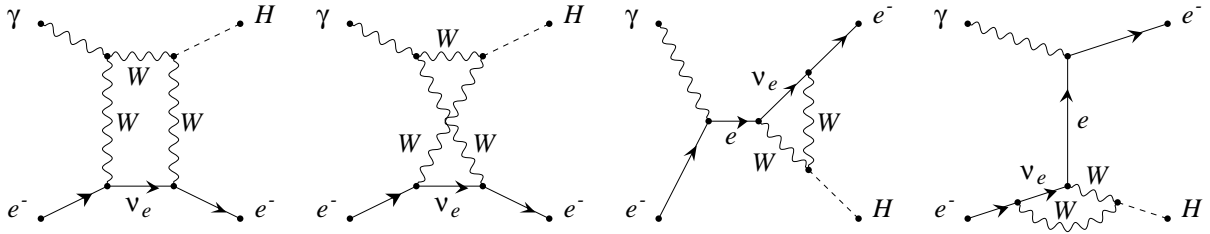} }
\end{picture} 
\caption{Subset of Feynman diagrams with $W$-box loop and related $eeH$ vertex.}
\label{dgrm3}  
\end{figure}

\begin{figure}
\begin{picture}(160,100)
\put(0,-70){\epsfxsize=20cm \leavevmode \epsfbox{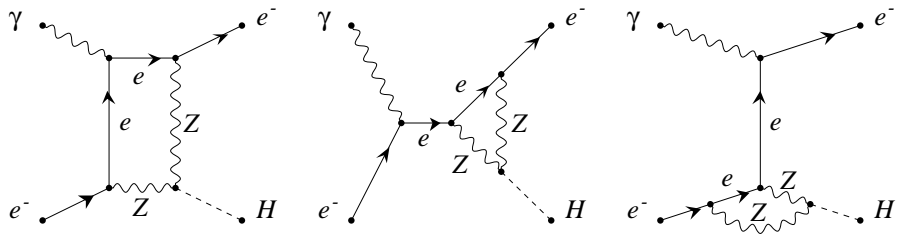} }
\end{picture} 
\caption{Subset of diagrams with $Z$-box loop and related $eeH$ vertex.}
\label{dgrm4}  
\end{figure}

In the {\it unitary gauge}, the Feynman diagrams 
 which contribute to this process are 
given in figures~\ref{dgrm1}--\ref{dgrm4} 
(for the figures of the Feynman diagrams, we used the program GRACEFIG
created by S.~Kawabata).
In figures~\ref{dgrm1} and \ref{dgrm2}, we show the fermion and $W$  
triangle loop respectively,  
with both $\gamma$ and $Z$ exchange in the $t$-channel,
where in the fermion loop we consider only the contribution
 of the top quark.
To these diagrams, the corresponding ones 
with opposite orientation for the fermion and $W$ loop have to be added.
In figures~\ref{dgrm3} and \ref{dgrm4}, the $W$-box and $Z$-box,
along with the related $eeH$ vertex diagrams, are presented, respectively.
The $eeH$ vertex cannot be neglected in the chiral limit, 
since only the divergent part of this vertex is proportional 
to the electron mass. Indeed, its finite part  
is proportional to the momentum square of the off-shell 
electron, and it is not zero in the chiral limit. 
Moreover, the finite part of the $eeH$ vertex is needed 
for the gauge invariance of the total amplitude.

In the 't-Hooft-Feynman gauge, we have to add to the first diagram in 
figure~\ref{dgrm2} the one where the $W$-lines are substituted
by the $W$-ghosts. Furthermore, we have to add to the diagrams of 
figures~\ref{dgrm2} and \ref{dgrm3} the ones where 
the $W$-lines in the loops are substituted by different combinations of
$W$-boson and $W$-goldstone propagators. For example, there are two diagrams
associated with the box diagrams in figure~\ref{dgrm3},
 where the $W$ propagator not connected with the electron is 
substituted by a $W$-goldstone propagator.
However, starting from the topology of the first two diagrams 
in figure~\ref{dgrm2}, 
there are new diagrams to add that cannot be generated by
the above rule.
The latter contain 4-legs vertices where the photon interacts 
with a $W$-goldstone and a Higgs boson, both in the $\gamma$ and $Z\;$ 
$t$-channel.
In particular, in the 't-Hooft-Feynman gauge, 52 diagrams replace
the first two of figure \ref{dgrm2}: 26 with 
$\gamma$-exchange plus 26  with $Z$-exchange in $t$-channel.
In the following, when we refer to figures~\ref{dgrm2} and \ref{dgrm3},
the complete subsets of the corresponding diagrams in
the 't-Hooft-Feynman gauge are implied.

The third diagram of figure~\ref{dgrm2} is given by the insertion of 
the vertex $Z\gamma H$ proportional to the counterterm coming from
the renormalization of the $Z$-$\gamma$ mixing self-energy function
at the one-loop level. This diagram is necessary to 
provide the ultra-violet finiteness of the W-loop contributions.
In our calculation, we have used the on-shell renormalization scheme. 
Hence, explicit contributions from the diagrams with
self-energy functions are missing.

\begin{figure}
\begin{picture}(160,100)
\put(-40,-70){\epsfxsize=20cm \leavevmode \epsfbox{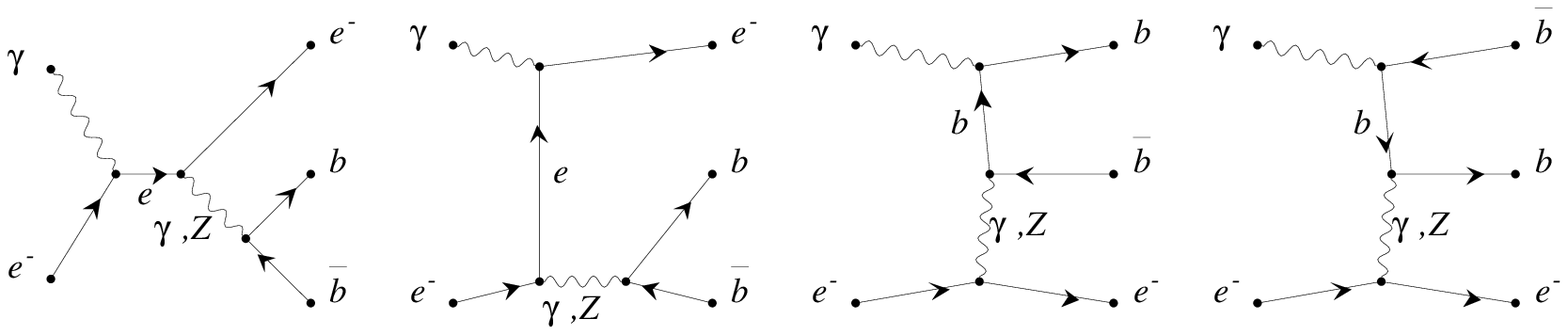} }
\end{picture} 
\caption{Diagrams for the \eaebb background.}
\label{dgrm5}  
\end{figure}

The total amplitude for the process (\ref{proc}) is of course
QED gauge invariant.
This means that, replacing in each diagram the photon polarization vector  
$e^{\mu}(\lambda,k_2)$ by its momentum $k_2^{\mu}$ (here and below 
$\lambda\equiv\pg =\pm 1$ is the photon
helicity), the sum over the whole set of diagrams has to vanish.
In general, a single diagram (or subset of diagrams)
is not transverse in the photon momentum by itself, but
we can select the non-transverse part
by just taking the terms  that do not vanish  
after this substitution.
Then, we find that it is possible to divide the whole set of 
diagrams into a few QED gauge--invariant subsets of diagrams :

\begin{itemize}
\item[i)] fermion--loops with $\gamma$ in the $t$-channel (figure~\ref{dgrm1});
\item[ii)] fermion--loops with $Z$ in the $t$-channel (figure~\ref{dgrm1});
\item[iii)] 
$W$ triangle--loops (figure~\ref{dgrm2}) + $W$ boxes and related $eeH$ 
vertices
(figure~\ref{dgrm3});
\item[iv)] 
$Z$--box and related $eeH$ vertices (figure~\ref{dgrm4}).
\end{itemize}

Below, we will see that the $W$-triangle and $W$-box diagrams
give rise to QED gauge non--invariant terms that cancel in the sum.

In order to get the analytical expression for the amplitude 
as a function of the
initial particle helicities, we decompose the 
Feynman amplitude in terms of the so-called {\it standard matrix
elements} defined as
\bea
 M_1(\sigma,\lambda) &\equiv & \phantom{-}
      \bar u^\sigma(k_3){\hat e}(\lambda) u^\sigma(k_1)\;, \nonumber\\
 M_2(\sigma,\lambda) &\equiv &\phantom{-}
      \bar u^\sigma(k_3){\hat k}_2 u^\sigma(k_1)\cdot (e(\lambda,k_2),k_3)\;, 
  \label{SME}    \\
 M_3(\sigma,\lambda) &\equiv & 
      -\bar u^\sigma(k_3){\hat k}_2 u^\sigma(k_1)\cdot (e(\lambda,k_2),k_1).
  \nonumber
\eea
Here $u^\sigma(k)$ denotes a spinor state for electrons with helicity 
$\sigma/2$ ($\sigma\equiv\pe =\pm 1$) and momentum $k$ 
($\hat k = k^\mu \gamma_\mu$, where  
$\gamma_\mu$ are the  Dirac's $\gamma$-matrices). Note that these elements 
contain the complete information about the polarizations of the initial 
electrons and photons.

\noindent
The amplitude can be expressed in terms of the standard matrix elements 
in the center-of-mass system (CMS), where the latter 
become\footnote{The same expressions still hold after a Lorentz boost 
along the collision axis of the process \eaeh.}:
\beq
   M_1(\sigma,\lambda) 
          \,=\,-\sqrt{-\frac{t}{2}}\,(1+\sigma\lambda),    \qquad 
   M_2(\sigma,\lambda) 
          \,=\,\sqrt{-\frac{t}{2}}\,u \; ,  \qquad
   M_3(\sigma,\lambda) \,=\,0 \; 
\label{resMi}
 \eeq 
with the Mandelstam variables defined as:
\beq
s=(k_1+k_2)^2,~~~t=(k_1-k_3)^2,~~~u=(k_2-k_4)^2.
\eeq
The last equality in (\ref{resMi}) is due to the orthogonality in the CMS 
of the photon polarization vector $e(\lambda,k_2)$ and the 
electron momentum $k_1$.
Nevertheless, we calculate the coefficients of
$M_3$, too. These coefficients
will be useful for the analysis of the QED gauge invariance of the result. 
Moreover, they will help us to get more compact analytical answers.

The amplitude corresponding to each diagram can be expressed in the following 
form:
\beq
M_1(\sigma,\lambda) \cdot {\cal B}_1 \;+\;
  M_2(\sigma,\lambda) \cdot {\cal B}_2 \;+\;
  M_3(\sigma,\lambda) \cdot {\cal B}_3\, ,
\label{bidef}
\eeq
where the coefficients ${\cal B}_i$ include the loop integrals.
The transversality in the photon momentum implies a linear relation between the 
coefficients ${\cal B}_i \;$ for the QED gauge--invariant sector of each diagram or subset of diagrams.
In fact, after the substitution 
$e^{\mu}\rightarrow k_2^{\mu}$ in the standard matrix elements 
(\ref{SME}) and then in (\ref{bidef}), we find that, 
for each QED gauge--invariant subset in the amplitude, 
these coefficients satisfy the following identity:
\beq
{\cal B}_1-\frac{u}{2}{\cal B}_2-\frac{s}{2}{\cal B}_3=0.
\label{qedidentity}
\eeq
As a consequence, the partial amplitudes can be represented in the
 following form, where the standard elements are substituted by their explicit
 values (\ref{resMi}):
\beq
  \frac{1}{2} \sqrt{-\frac{t}{2}} \;
  [(u{\cal B}_2-s{\cal B}_3) -\sigma\lambda(u{\cal B}_2+s{\cal B}_3)]\;.
  \label{B23form}
\eeq

The final result can be further simplified
by using the crossing symmetry connecting the process (\ref{proc}) 
to the crossed one $e^+(k_3) \gamma(k_2) \rightarrow e^+(k_1) H(k_4)$. 
If we perform a crossing transformation
$(s\leftrightarrow u,\; \sigma\to -\sigma)$ 
in the matrix element of the process (\ref{proc}),
we get the same function with opposite electric charge and $Z$-charge
of the electron,(i.e., with $Q_e \to -Q_e$, and
$g_e \to -g_e$). Of course, for the $Z$-charge
the change of the electron helicity has to be taken into account, too.
We found that, for the QED gauge--invariant component of 
each subset of the diagrams
represented in figures~\ref{dgrm1}--\ref{dgrm4}, this symmetry is
fulfilled, what simplifies further our formulas.
Indeed, this symmetry implies that the form factor $u{\cal B}_2-s{\cal B}_3$ is
antisymmetric with respect to the substitution $s\leftrightarrow u$, while
$u{\cal B}_2+s{\cal B}_3$ is symmetric (i.e.,
${\cal B}_2\leftrightarrow {\cal B}_3$ for $s\leftrightarrow u$).
Then, instead of the coefficients ${\cal B}_2$ and ${\cal B}_3$, 
it is worthwhile to consider
the following symmetric and antisymmetric combinations:
\beq
 {\cal F}^s \;\equiv\; \frac{u{\cal B}_2+s{\cal B}_3}{2}\;, \qquad
   {\cal F}^a \;\equiv\; \frac{u{\cal B}_2-s{\cal B}_3}{2}\;.
\label{Ffactors}
\eeq
Using the above form factors,  our final analytical results
can be represented in the following compact form
\beq
\sqrt{-\frac{t}{2}} \; [{\cal F}^a \,-\,\sigma\lambda{\cal F}^s ] \;.
\label{Fasform}
\eeq
Note that all the dependence on the photon helicity is concentrated 
in the explicit
factor of the second term in (\ref{Fasform}). 
Hence, when one averages over the photon helicity,
the dependence on the electron helicity
arises only from the $Ze\bar e$ coupling. Since 
$g_e^-\simeq -0.658$ and
$g_e^+\simeq 0.538$, the  $Z\;$ $t$-channel contributions 
have opposite signs in the
amplitudes for left-handed and right-handed electrons. 
Moreover, in the region of large transverse momentum, there is a moderate 
difference between the $\gamma$ and $Z\;$ $t$-channel propagators.
This explains the  destructive interference of the $\gamma$ and 
$Z\;$ $t$-channels for a right-handed electron beam, and the 
mutual enhancement of these contributions in the 
case of left-handed electrons,  when the photon beam is unpolarized
(see also section 5).

Because of the crossing symmetry for $s\leftrightarrow u$,
one has  ${\cal B}_2={\cal B}_3$ in eq. (\ref{bidef}),
for all the triangle amplitudes.
This means that, for this class of diagrams, the formula (\ref{Fasform}) can
be further simplified into:
\begin{equation}
\frac{1}{2} \; \Lambda(\sigma,\lambda) \cdot {\cal T}\; ,
\label{Tr-form}
\end{equation}
where ${\cal T}={\cal B}_2={\cal B}_3$ and
\begin{equation}
\Lambda(\sigma,\lambda) \;\equiv\; 
   \sqrt{-\frac{t}{2}}\;[(u-s)\,-\,\sigma\lambda(u+s)] \;.
\label{Lambda}
\end{equation}
Finally,
the differential cross section for the process with longitudinally 
polarized photon and electron beams 
\footnote{Of course, the unpolarized cross section can be obtained by 
averaging over the helicity of the initial particle(s).}
is given by
\begin{equation}
 \frac{d\sigma(e\gamma\to eH)}{d\Omega} \;=\;
   \frac{1-m_H^2/s}{64\pi^2s} \; (\alpha^2 M_Z)^2\; |M(\sigma,\lambda))|^2,
\end{equation}
where $\Omega$ is the spherical scattering angle of the final electron,
and
the analytical expression for the total QED gauge--invariant matrix element 
can be expressed as:
\bea
{\cal M}(\sigma,\lambda)&=&
{\cal M}_{\triangle_\gamma^f}(\sigma,\lambda)+
{\cal M}_{\triangle_Z^f}(\sigma,\lambda)+
{\cal M}_{\triangle_{\gamma,1}^W}(\sigma,\lambda)+
{\cal M}_{\triangle_{\gamma,2}^W}(\sigma,\lambda)+
\nonumber\\
&&
{\cal M}_{\triangle_{Z,1}^W}(\sigma,\lambda)+
{\cal M}_{\triangle_{Z,2}^W}(\sigma,\lambda)+
{\cal M}_{\Box_W}(\sigma,\lambda)+
{\cal M}_{\Box_Z}(\sigma,\lambda)\, .
\label{Ampl}
\eea
The partial amplitudes ${\cal M}_{i}$ are given by
the following QED gauge--invariant contributions:
\begin{itemize}
\item Triangle fermion loops:
\end{itemize}
\beq
 {\cal M}_{\triangle_{(\gamma,Z)}^f}(\sigma,\lambda) \,=\,
     \frac{m_t^2}{M_Z^2}\;N_c
     \frac{Q_t}{s_w c_w} \;{\cal P}_{(\gamma,Z)}^f
     \cdot \Lambda(\sigma,\lambda)\cdot
     \left[{\cal T}_1(m_t) -2{\cal T}_2(m_t)\right]
\label{triferm}
\eeq
$$
 {\cal P}_{\gamma}^f \,=\, \frac{2  Q_e Q_t}{-t}\, , \qquad 
 {\cal P}_{Z}^f \,=\, \frac{g_e^\sigma (g_t^+ + g_t^-)}{-t+M_Z^2}\, .
$$
Here, ${\cal M}_{\triangle_\gamma^f}$ and ${\cal M}_{\triangle_Z^f}$
represent the contributions of the $\gamma$ and $Z\;$ 
$t$-channel, respectively.
In the above formulae, $N_c=3$ is the color weight of the $t$-quark, 
while the electric and $Z$ charges of the fermions, $Q_f$ and $g_f^{\pm}$, 
are given by
$$
Q_e = -1, ~~~
g_e^{+}=-Q_e\frac{s_w}{c_w},~~~
g_e^{-}=-\frac{1/2+Q_e s_w^2}{s_wc_w},
$$
$$
Q_t = \frac{2}{3}, ~~~
g_t^{+}=-Q_t\frac{s_w}{c_w},~~~
g_t^{-}=\frac{1/2-Q_t s_w^2}{s_wc_w}.
$$
Also, $s_w \equiv \sin \theta_W$ and $c_w \equiv \cos \theta_W$,
with $\theta_W$ the Weinberg angle.

The expressions for the one-loop form factors ${\cal T}_1(m)$ 
and ${\cal T}_2(m)$ are given by:
\bea
{\cal T}_1(m) &\equiv & C_0(m,-k_4,m,k_3-k_1,m),
\nonumber \\
{\cal T}_2(m) &\equiv & \frac{1}{s+u} \left\{ 2m^2 \cdot {\cal T}_1(m)+
  \frac{t}{s+u}\cdot [B_0(t,m^2,m^2)-B_0(m_H^2,m^2,m^2)] \;-\,1 \right\}
\label{tfunc}
\eea
where the functions $C_0$ and $B_0$ are defined in the 
Appendix\footnote{Note that,
in the limit $t\to 0$, the $B_0$ integrals do not contribute due to the 
factor $t$ in (\ref{tfunc}). Hence, the $C_0$ integrals give 
the dominant contribution to the total cross section.}.
The form factors ${\cal T}_1(m)$ and ${\cal T}_2(m)$
can also be expressed in terms of elementary functions (see
\cite{haau,haad}).

\begin{itemize}
\item Triangle W loops:
\end{itemize}
\beq
 {\cal M}_{\triangle_{(\gamma,Z),1}^W}(\sigma,\lambda) \;=\;
      \frac{m_H^2}{M_Z^2}\;
     \frac{1}{s_w c_w} \;{\cal P}_{(\gamma,Z),1}^W 
     \cdot \Lambda(\sigma,\lambda)\cdot
     \left[{\cal T}_2(M_W)\right]\;,                 
\label{triwf}
\eeq
$$
 {\cal P}_{\gamma,1}^W \,=\, \frac{Q_e}{-t}\, , \qquad 
 {\cal P}_{Z,1}^W \,=\, \frac{g_e^\sigma (1-2s_w^2)}{2s_w c_w}
                       \cdot  \frac{1}{-t+M_Z^2}\, ,
$$
The terms ${\cal M}_{\triangle_{(\gamma,Z),1}^W}$
include the loop with the $W$-goldstone propagators only. Three diagrams
contribute both in the photon and Z exchange. The presence of 
 terms proportional to $m_H^2/M_Z^2$ 
is a consequence of the Higgs mechanism and the  decoupling of
the longitudinal components of the $W$ boson at high energy.

\beq
  {\cal M}_{\triangle_{(\gamma,Z),2}^W}(\sigma,\lambda) \;=\;
  \frac{c_w}{s_w} \; {\cal P}_{(\gamma,Z),2} 
  \cdot \Lambda(\sigma,\lambda)\cdot
  \left[- A^{(\gamma,Z)}_1 {\cal T}_1(M_W)
     +A^{(\gamma,Z)}_2 {\cal T}_2(M_W) \right],
\eeq
$$
 {\cal P}_{\gamma,2}^W \,=\, \frac{Q_e}{-t}\, , \qquad 
 A^\gamma_1 \,=\, 8, \qquad A^\gamma_2 \,=\, 6, 
$$
$$
 {\cal P}_{Z,2}^W \,=\, \frac{g_e^\sigma}{-t+M_Z^2}\, ,\qquad
  A^Z_1 \,=\,6\frac{c_w}{s_w}-2\frac{s_w}{c_w}\, , \qquad 
  A^Z_2 \,=\, 5\frac{c_w}{s_w}-\frac{s_w}{c_w}. 
$$
The terms ${\cal M}_{\triangle_{(\gamma,Z),2}^W}$
include the contribution of the $W$-triangle diagrams 
with $W$-ghosts and with a 
mixture of $W$-bosons and $W$-goldstones running in the loop.
In the ${\cal M}_{\triangle_{Z,2}^W}$ term,
we also include the diagram with the $Z\gamma H$ counterterm. 

\begin{itemize}
\item $W$ and $Z$ boxes with related $eeH$ vertices:
\end{itemize}
\beq
 {\cal M}_{\Box_{(W,Z)}}(\sigma,\lambda) \;=\; 
     \frac{{\cal P}_{(W,Z)}^\Box}{s_w c_w}  \;
     \sqrt{-\frac{t}{2}} \cdot
     \left[{\cal D}_{(W,Z)}^a - \sigma\lambda{\cal D}_{(W,Z)}^s \right]\, ,
\label{box}
\eeq
$$
 {\cal P}_{W}^\Box = -2\frac{c_w^2}{s_w^2} \delta_e^\sigma, \qquad
 {\cal P}_{Z}^\Box = 4(g_e^\sigma)^2 Q_e, $$
where $\delta_e^{+}=0,\; \delta_e^{-}=1$, and the form 
factors $D_{(W,Z)}^{s,a}$ 
are the symmetric and antisymmetric parts of the 
functions
\bea
 {\cal D}_W \;&\equiv\;& 
  u(D^W_{3}+D^W_{23}+{\tilde D}^W_1-{\tilde D}^W_{12}-{\tilde D}^W_{22})\, ,
\label{wzbox}
\\
{\cal D}_Z \;&\equiv\;& u(D^Z_{12}+D^Z_{22})
\nonumber
\eea
under the crossing $s\leftrightarrow u$ symmetry transformation.
The functions $D^{(W,Z)}_i$ appearing in 
eqs.~(\ref{wzbox}) contain the results of the box loop-integrals,
and are related to the integrals
defined in the Appendix in the following way:
\bea
D^W_i &=& D_i(0,k_1,M_W,k_1+k_2,M_W,k_3,M_W), \quad 
   \tilde{D}^{W}_i=D^{W}_i|_{s\leftrightarrow u},
   \label{fwbox}\\
D^Z_i &=& D_i(m_e,k_2,m_e,k_2-k_3,M_Z,-k_1,M_Z).
\label{fzbox}
\eea
In the $Z$-box functions in eq.~(\ref{fzbox}), we restore the
electron mass since the integrals $D_i$ are not separately finite for $m_e=0$. 
Of course, 
the total amplitude has to be insensitive to the value 
of $m_e$ used to regularize each singular $Z$-box integral.
We checked the stability of the total result  numerically, 
for a wide range of $m_e$, going from 
$10^{-34}$GeV up to its physical value.

In the following sections,
 we refer to the different QED gauge--invariant contributions
defined above as:
\begin{center}
\begin{tabular}{lll}
 "$\gamma\gamma H$" &corresponding to& ${\cal M}_{\triangle_\gamma^f}
                        +{\cal M}_{\triangle_{\gamma,1}^W}
                         +{\cal M}_{\triangle_{\gamma,2}^W}$ \\
 "$Z\gamma H$"      &corresponding to& ${\cal M}_{\triangle_Z^f}
                         +{\cal M}_{\triangle_{Z,1}^W}
                         +{\cal M}_{\triangle_{Z,2}^W}$ \\
  "$BOX$"  &corresponding to&    ${\cal M}_{\Box_W}+{\cal M}_{\Box_Z}$
\end{tabular}
\end{center}

For completeness, we now show the analytical results for 
the QED gauge non--invariant terms,
arising from the $W$-triangle and $W$-box diagrams.
As we checked, the sum of these terms vanishes.
They can be worked out by calculating all the 
coefficients ${\cal B}_i$ and then performing the substitutions 
$e(\lambda,k_2)\to k_2$ in (\ref{bidef}). From the $W$-triangle diagrams, 
we obtain for such terms
\beq
{\cal M}_{\triangle_{(\gamma,Z),3}^W}(\sigma,\lambda) \,=\,
     \frac{c_w}{s_w} \; {\cal P}_{(\gamma,Z),3}
     \cdot [{\cal K}_{(\gamma,Z)} \cdot
     M_1(\sigma,\lambda)\cdot {\cal T}_1(M_W)],
\label{nig1}
\eeq
$$
  {\cal P}_{\gamma,3} \;=\; -\frac{Q_e}{-t}, \qquad
  {\cal P}_{Z,3}      \;=\; \frac{c_w g_e^\sigma}{s_w}\cdot \frac{1}{-t+M_Z^2}.
$$
In eq.~(\ref{nig1}), the coefficients ${\cal K}_{(\gamma,Z)}$, 
arising from the W-triangle diagrams with W-ghosts and with a
mixture of W-bosons and W-goldstones running in the loop,
are given by
\beq
{\cal K}_\gamma \;=\; 3t\, , \qquad {\cal K}_Z \;=\; 3(-t+M_Z^2) \,,
\eeq
and cancel the corresponding $\gamma$ and $Z$ propagators.
After summing the two terms in (\ref{nig1}), we obtain
\beq
 {\cal M}_{\triangle_{(\gamma+Z),3}^W}(\sigma,\lambda) \,=\, 
     \frac{c_w\delta_e^\sigma}{2 s_w^3} \;[-3 M_1(\sigma,\lambda) 
     \cdot {\cal T}_1(M_W)]. 
\label{T-noninv}
\eeq
This term is exactly cancelled by an opposite term coming 
from the QED gauge non--invariant contribution of
the $W$-box diagrams (figure~\ref{dgrm3}).
\\
We stress that the $eeH$-vertex diagrams must be  added to
the $W$-box diagrams in order to fulfil the QED gauge--invariance 
identity (\ref{qedidentity}) for this subset of diagrams, after the
cancellation of the term (\ref{T-noninv}). Note that the $eeH$-vertex diagrams
contribute through the ${\cal B}_1$ coefficients to the  weight
of the standard matrix
element $M_1$ (\ref{SME}). 
In the case of the $Z$-box diagram (figure~\ref{dgrm4}),
the identity (\ref{qedidentity}) is fulfilled automatically when one adds
the corresponding $eeH$ vertex diagrams.

We checked that our expressions for the one-loop form factors 
agree with that of ref. \cite{barr,abba,djou}, where the
crossed process $e^+e^-\rightarrow \gamma H$ was investigated.
We also checked that our loop form factors ${\cal B}_i$ 
are in agreement
with the corresponding loop form factors for the process $e^+e^-\to H Z$ in
\cite{denu}, if the  proper crossing transformation is made and the
$Z$ vertices are replaced by the related $\gamma$ vertices.

Some comments on the gauge dependence of our
decomposition in $\gamma\gamma H$, $Z\gamma H$
and $BOX$ contributions of eq.~(\ref{Ampl}) are in order.
By construction, the identity (6), which is equivalent to the photon
transversality, is fulfilled by these contributions
separately.
One can check that our decomposition corresponds to the
$\gamma\gamma H$, $Z\gamma H$ and $e^+e^-\gamma H$ Green functions in the 
so-called non-linear gauge, where the derivative in the 't-Hooft-Feynman gauge
for the $W$ field is replaced by the corresponding covariant derivative. 
It has been  shown \cite{barr} that, in this gauge,  the 
Slavnov-Taylor identities for the $\gamma\gamma H$ and $Z\gamma H$ Green
functions are simply equivalent to the transversality with respect to 
the photon momenta.

Here, we want to stress, first of all, some technical advantages of 
the proposed calculation method, 
based on exploiting the transversality identity (6).
Indeed, we chose the widely used 't-Hooft-Feynman gauge, 
and decomposed the Feynman amplitudes in terms of
a set of standard matrix elements. We used the set (2), although
this choice is not unique. 
Then, using the transversality of the physical amplitudes, 
we found that the coefficients of the standard
matrix elements have to fulfill some linear identity, that in our case is
eq. (6). Accordingly, one can take
any partial contribution, for example the contribution of different
subsets of diagrams, then calculate only the terms satisfying this
linear identity, and ignore the violating terms.
As a consequence, as we showed, this technique helps getting
more compact answers for the physical amplitudes. 
We stress the generality of the proposed technique. 
For comparison, we refer to the paper \cite{djou}, 
where the same (QED gauge invariant) contributions
to the crossed process $e^+e^-\to \gamma H$ were obtained by 
choosing an {\it ad hoc} special set of standard matrix elements.

 There is another advantage of the adopted decomposition.  
Our main goal here is to demonstrate  the usefulness of the
process $e\gamma\to eH$ for measuring the $Z\gamma H$ coupling.
Of course, one does not expect that the contribution of the 
$Z$-boson vertices can be separated from the related photon vertices in a
SU(2) gauge invariant way in the standard model. 
This connects
with the presence of the third component of the SU(2) gauge field in both 
the $Z$-boson and photon fields. 
Moreover, one can show that even the box 
contribution can not be isolated in a SU(2) gauge invariant way,
by comparing the results in  
the linear 't-Hooft-Feynman gauge used here, with the
nonlinear gauge results  (see details in \cite{barr}). 
However, we will see in the next section that the contribution of 
the  box diagrams to the cross section is in general rather small.
Hence, the most important issue is the separation of the \zah and \aah
vertices.
In general, if one wants to compare  the relative contributions of the 
\zah and \aah vertices, one needs to specify the gauge in which one works.

Note, that it is possible that there are new non-standard
particles circulating in  the loops of 
the  \zah and \aah Green functions,
giving  additional contributions to the ones of 
the standard-model electroweak theory. 
Hence, measuring the corresponding amplitudes 
could give us some hints on the nature of the actual extension of  
the standard model. 

In case the new scenario implies the non-decoupling regime for the Higgs boson
interaction with the new particles, 
we can use an effective point-like
interaction Lagrangian to calculate the contributions of the new physics. In
\cite{BW,H93}, the relevant lagrangian terms were 
classified and parameterized using five anomalous
coupling constants. Of course, these terms must be
$SU(2)\times U(1)$ gauge invariant, and, hence,
transversal with  respect to the photon momentum. The same is true for
their contributions to the \zah and \aah Green functions and 
to the cross section of the process under discussion. 
As a consequence, 
from a kinematical point of view, this type of new physics 
would contribute similarly to the QED gauge-invariant contributions of the 
standard \zah and \aah Green functions.
Thus, our strategy in the next sections will be to find out the kinematical
regions where the relative contribution of the transversal \zah Green
function is enhanced in comparison with the \aah one. 

Alternative cases, where the Higgs boson interacts with the new particles 
in the decoupling limit, can not be described by an effective point-like 
Lagrangian,  and some different strategy
 is necessary to  extract the contributions of the corresponding \zah and
\aah induced vertices (see \cite{holl}, for the case of the 
minimal supersymmetrical extension of the standard model). 
In these cases, there could be some kind
of common agreement to define the different contributions. For instance,
our decomposition in eq.(\ref{Ampl}) could make the job.
Consequently, the analysis of the  numerical results and kinematical cuts 
made in the following should be of some help for the 
measurement of the \zah induced vertex 
in decoupling cases, too.


\noindent
\section{Exact cross sections}
\vskip 10pt
In this section, we present the total rates $\sigma(He)$ for the process \eaeh
versus the Higgs boson mass $\mh$ and the c.m. \egam collision energy
$\sqs$. We also compare them with the cross sections for the competing 
tree-level process \eahnw \cite{eanh}. A possible strategy for enhancing
the \zah vertex effects with respect to the dominant \aah contribution
is then outlined.

In order to correctly relate our exact results to the previous approximate
estimates, one should take into account that in our paper we always 
assume an exactly monochromatic initial photon beam.
It has been customary for some time to present total rates convoluted with
a particular form of the initial-photon energy spectrum \cite{spec}.
On the other hand,  presenting
unfolded results can help in distinguishing the physical effects
related to the particular collision process from details 
depending on the final realization of the backscattered laser beam, that could 
evolve with time before the final project of the linear collider is approved
\footnote{That was  recently stressed by V.I.~Telnov \cite{teln}.}.

In our numerical results, we assume $\alpha(m_e)=1/137$ in 
each vertex that involves an on-shell (or almost on-shell) photon.
On the other hand, we express both purely electroweak vertices
and vertices involving off-shell photons (exchanged in the $t$ channel when
$\ptH \gta 10$GeV) in terms of $\alpha(\mw)=1/128$.
This is made in a gauge invariant way, i.e. by just rescaling the final 
cross sections.
Also,  we assumed $\mz=91.187$GeV, $\sin^2 \theta_W = 0.2247$, and, 
for the top-quark and $b$-quark masses, $m_t=175$GeV and 
$m_b=4.3$GeV, respectively.

\begin{figure*}[t]
\vspace*{-4.cm}
\begin{center}
 \mbox{\epsfxsize=16cm\epsfysize=18.5cm\epsffile{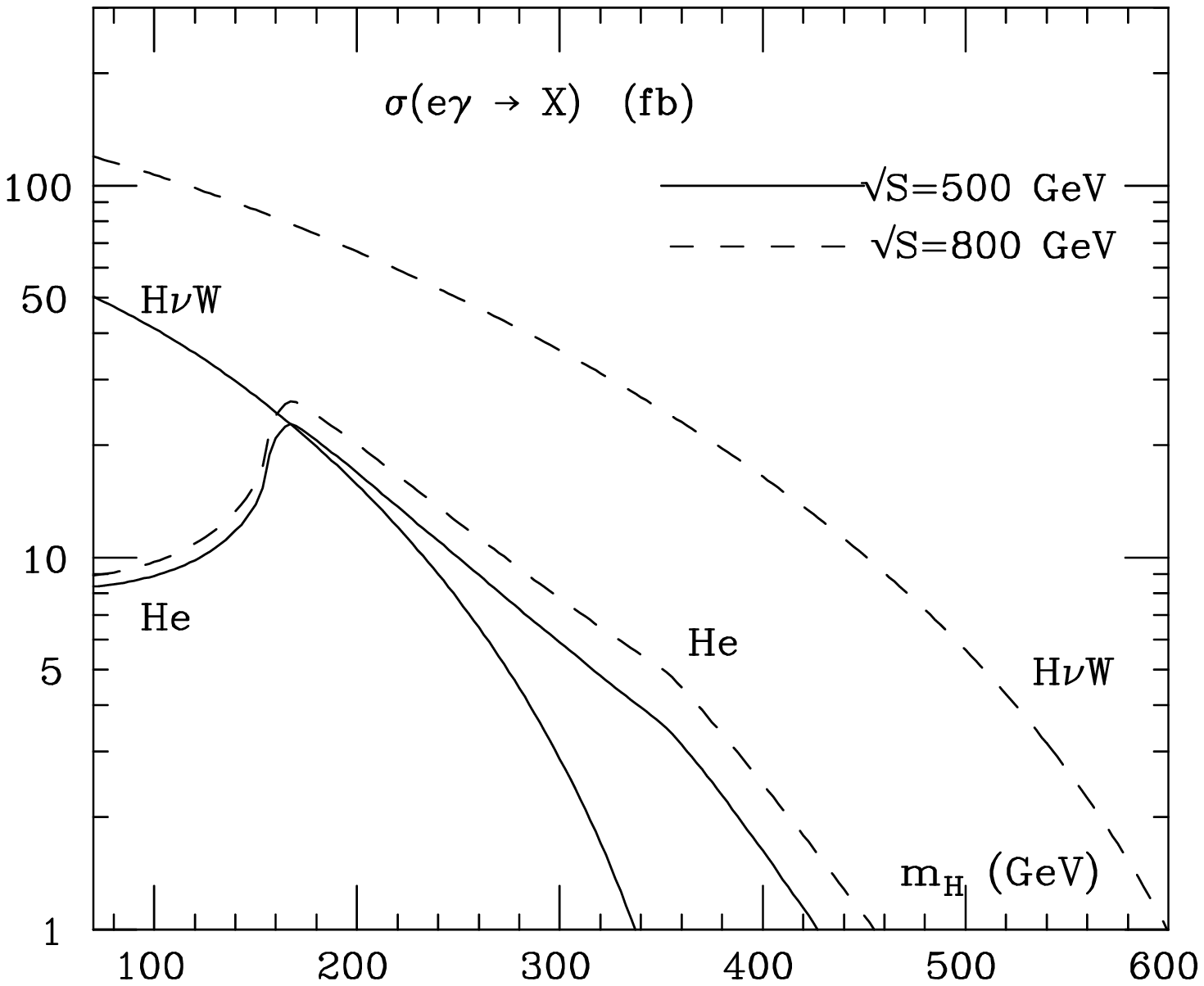}}
\vspace*{-5.5cm}
\caption{ Total cross sections for the two main H production processes.
 }
\label{fig31}
\end{center}
\vspace*{-4.5 cm}
\begin{center}
 \mbox{\epsfxsize=16cm\epsfysize=20cm\epsffile{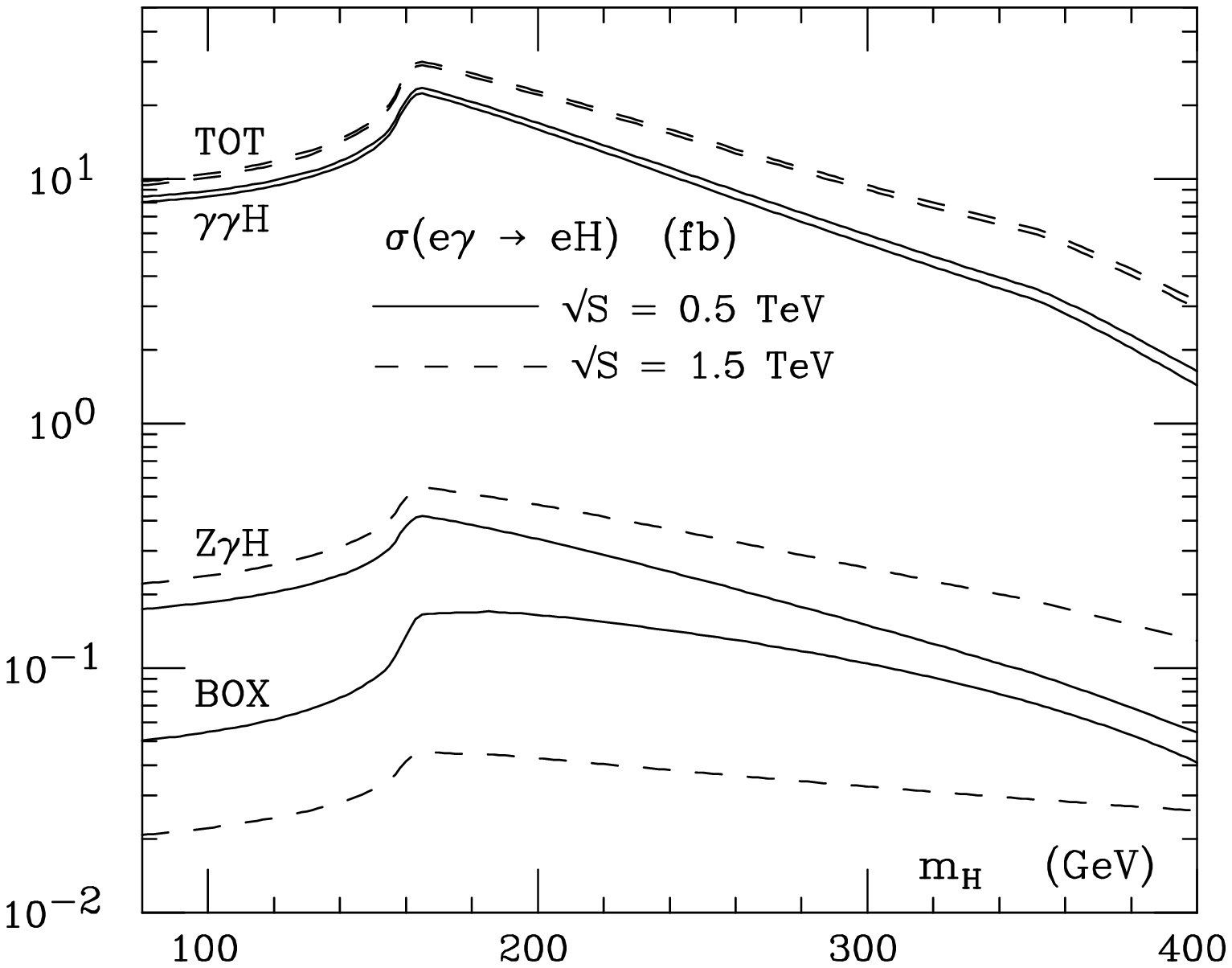}}
\vspace*{-6.cm}
\caption{ Total cross section for \eaeh plus different partial
contributions (see text).
 }
\label{fig33}
\end{center}
\end{figure*}

In figure \ref{fig31}, the total (unpolarized)
cross sections for the one-loop process
\eaeh (obtained by integrating the analytical formulae in section 2) 
and the tree-level Higgs production \eahnw (computed by CompHEP \cite{comp})
are plotted versus $\mh$, for $\sqs=500$ and 800 GeV.
Numerical results can also be found in table \ref{tab32}, where the
$\mh$ dependence of the two channels is reported for  
$\sqs=$0.5, 1 and 1.5 TeV.

One can see that the process \eaeh is characterized by relatively large
rates. For instance, for $\mh$ up to about 400 GeV, one finds
$\sigma(H e) >1$ fb, which, for an integrated luminosity of about
100fb$^{-1}$, corresponds to more than 100 events.
Note also that the cross section for the crossed process \eeah
has a similar behaviour with $\mh$, but is only about 
a fraction $(\frac{1}{200}\div\frac{1}{400})$  of $\sigma(H e)$, 
in the range $\mh=(100\div 400)$GeV at $\sqs=500$GeV \cite{djou}.
Moreover, contrary to $\sigma(H e)$, $\sigma(H \gamma)$ drops as $1/s$
at large c.m. collision energies .

At  $\sqs\simeq 500$ GeV, the \eaeh rate increases with $\mh$ up to
$\mh\simeq 2 \mw$, where  $\sigma(H e)\simeq 21$fb. 
For larger masses, the cross section
falls, but more slowly than in the $\sigma(H\nu W)$ case.
As a result, $\sigma(H e) > \sigma(H\nu W)$ for $\mh\gta 180$ GeV.

At larger $\sqs$, $\sigma(H e)$ increases, but only slightly. 
On the other hand, $\sigma(H \nu W)$ takes much advantage by a larger
c.m. collision energy and, e.g., at $\sqs=1$TeV and $\mh=180$GeV,
is more that a factor 4 larger than the corresponding 
\eaeh cross section (cf. table \ref{tab32}).

We also compared the exact rates for \eaeh with the rates one
obtains in the Weizs\"acker-Williams (WW) approximation according to the 
approach of \cite{ebol}.
We have found that the WW approximation differs from
the exact rate by less than 15\% in the range $\sqs=0.5-1.5$TeV,
working better at lower $\sqs$ and higher $\mh$.
For instance, at $\sqs=500$GeV and $\mh=300$GeV, the WW cross section
is larger than the exact one by only 3.4\%.
Anyhow, by adopting an improved WW approach \cite{frix},
one reaches an accuracy better than the 6\% in the same $\sqs$ range.  

Note also that the difference in the relative importance of the two channels
\eaeh and \eahnw $\;$ in  figure \ref{fig31} with respect 
to  figure 1 of \cite{ebol} is mainly due to the inclusion
of a spectrum for the photon beam \cite{spec} in the latter case.
Indeed, the photon spectrum considered 
depletes considerably $\sigma(H \nu W)$, while $\sigma(He)$ keeps
relatively stable.

In figure \ref{fig33},  we show separately the contributions
to the total cross section for \eaeh given by the squared amplitudes
corresponding to the subsets of Feynman diagrams ``\aah", ``\zah" and ``BOX"
(defined in section 2).
Even if this separation is by no means formally rigorous
(and neglects the relative interference effects), it can help 
in getting a feeling of the relative importance of triangular vertices and
box contributions to the total cross section.
In figure \ref{fig33}, the upper solid (dashed) curve corresponds
to the total cross section at $\sqs=0.5(1.5)$TeV.
The slightly lower curve shows the largely dominant contribution from
the \aah vertex graphs, while the \zah and BOX cross sections are
a factor about 50(45) and 150(500) smaller, respectively, 
for $\mh\sim 150$GeV and $\sqs=0.5(1.5)$TeV.
At larger $\mh$, this pattern keeps qualitatively similar.

In principle, the \eaeh total cross section 
(and its main contribution from \aah)
is of the same order of magnitude of the total rates  
for Higgs production in \gag collisions \cite{aahd}.
Indeed, the expected resolution on the beam energy
smears the higher peak cross section over a width
much larger than the Higgs resonance.
As a result, the channel \eaeh has a comparable potential with
respect to the
process $\gam\to H$ in testing the \aah vertex, as far as 
the production rates are concerned.
In this paper, on the other hand, we would like to concentrate
on the problem 
of disentangling the \zah vertex effects, which are out of the
\gag-collision domain. 

\begin{figure*}[t]
\vspace*{-3.5cm}
\begin{center}
 \mbox{\epsfxsize=11cm\epsfysize=14.cm\epsffile{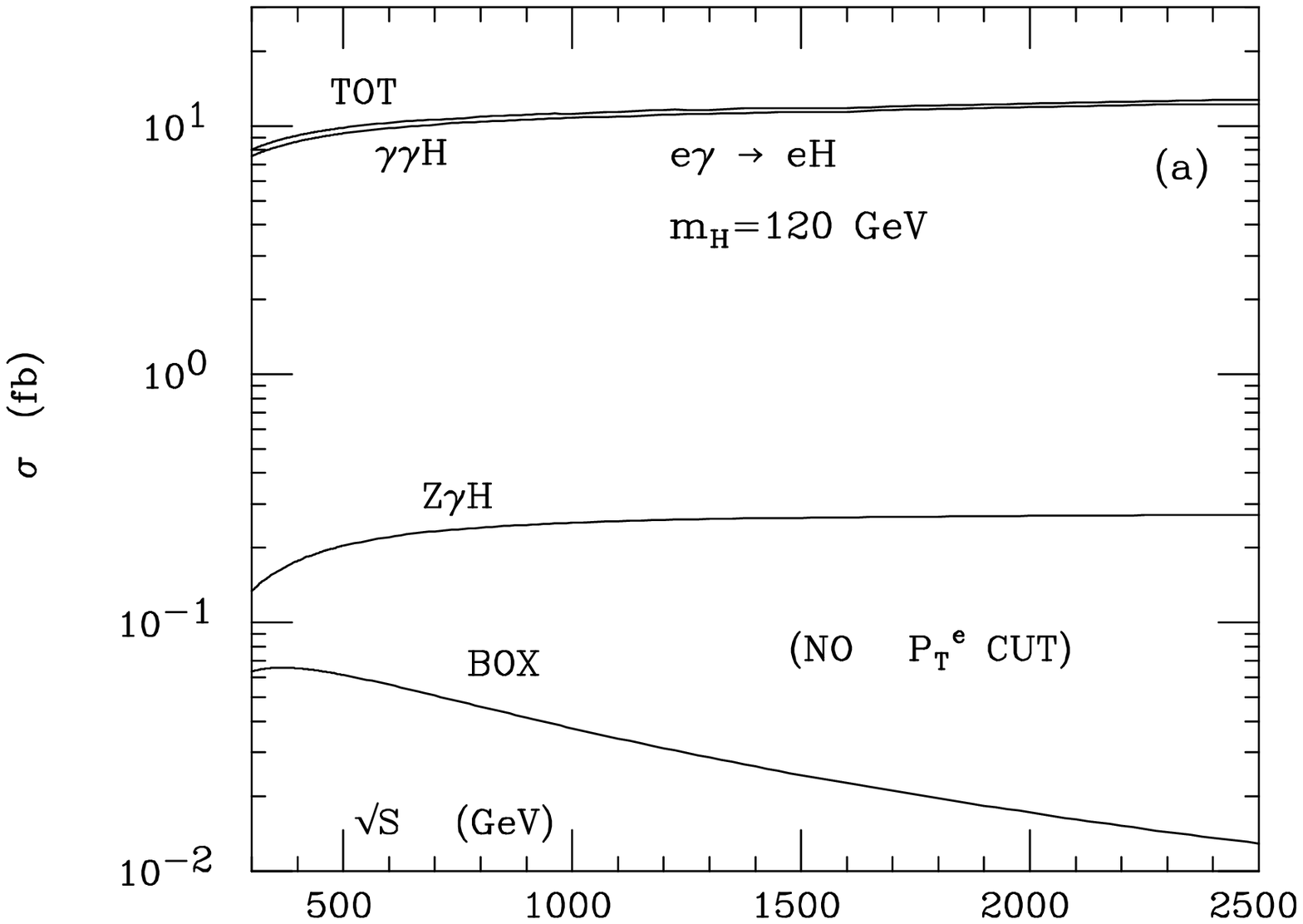}}
\end{center}
\vspace*{-8.cm}
\begin{center}
 \mbox{\epsfxsize=11cm\epsfysize=14.cm\epsffile{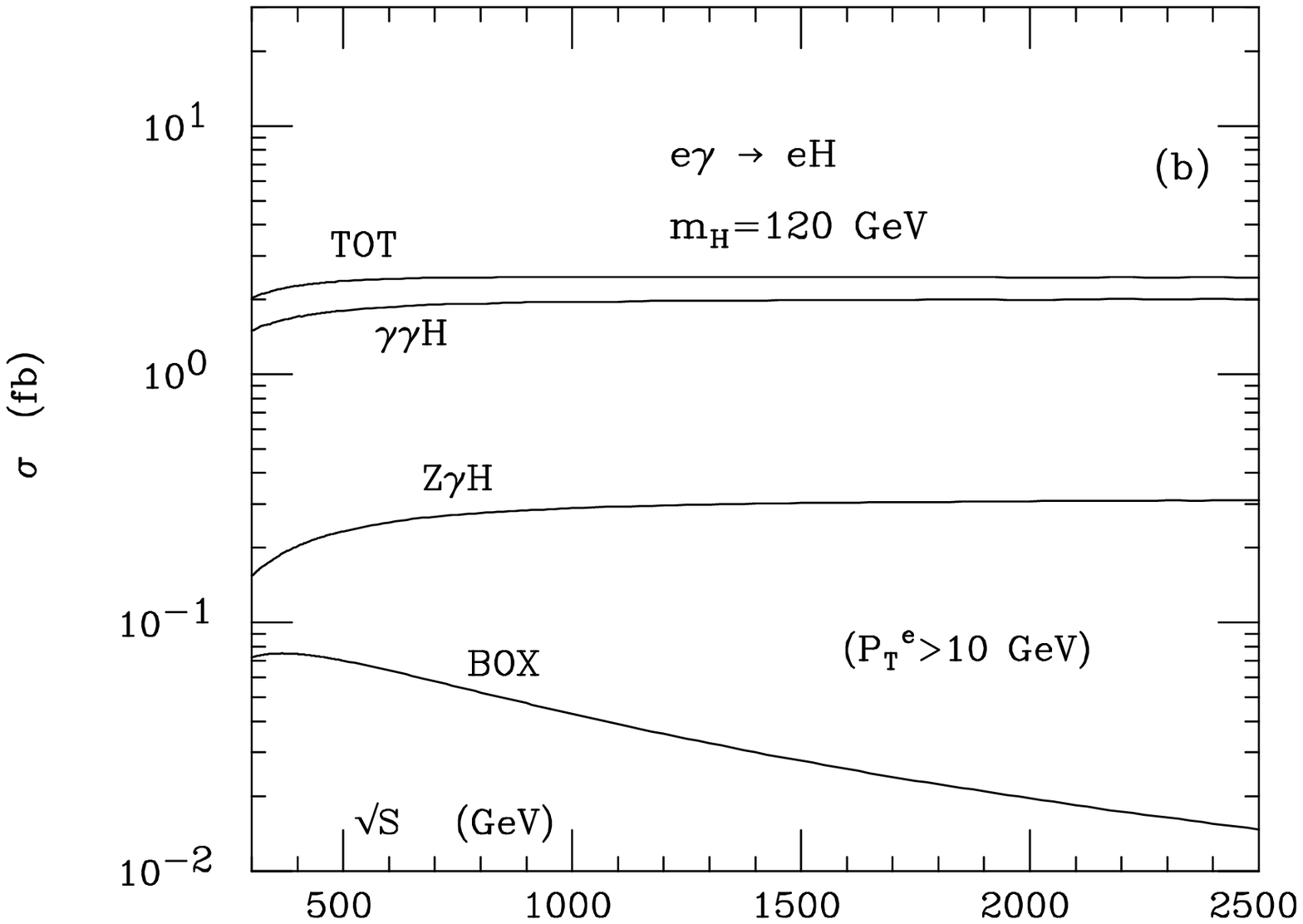}}
\end{center}
\vspace*{-8. cm}
\begin{center}
 \mbox{\epsfxsize=11cm\epsfysize=14.cm\epsffile{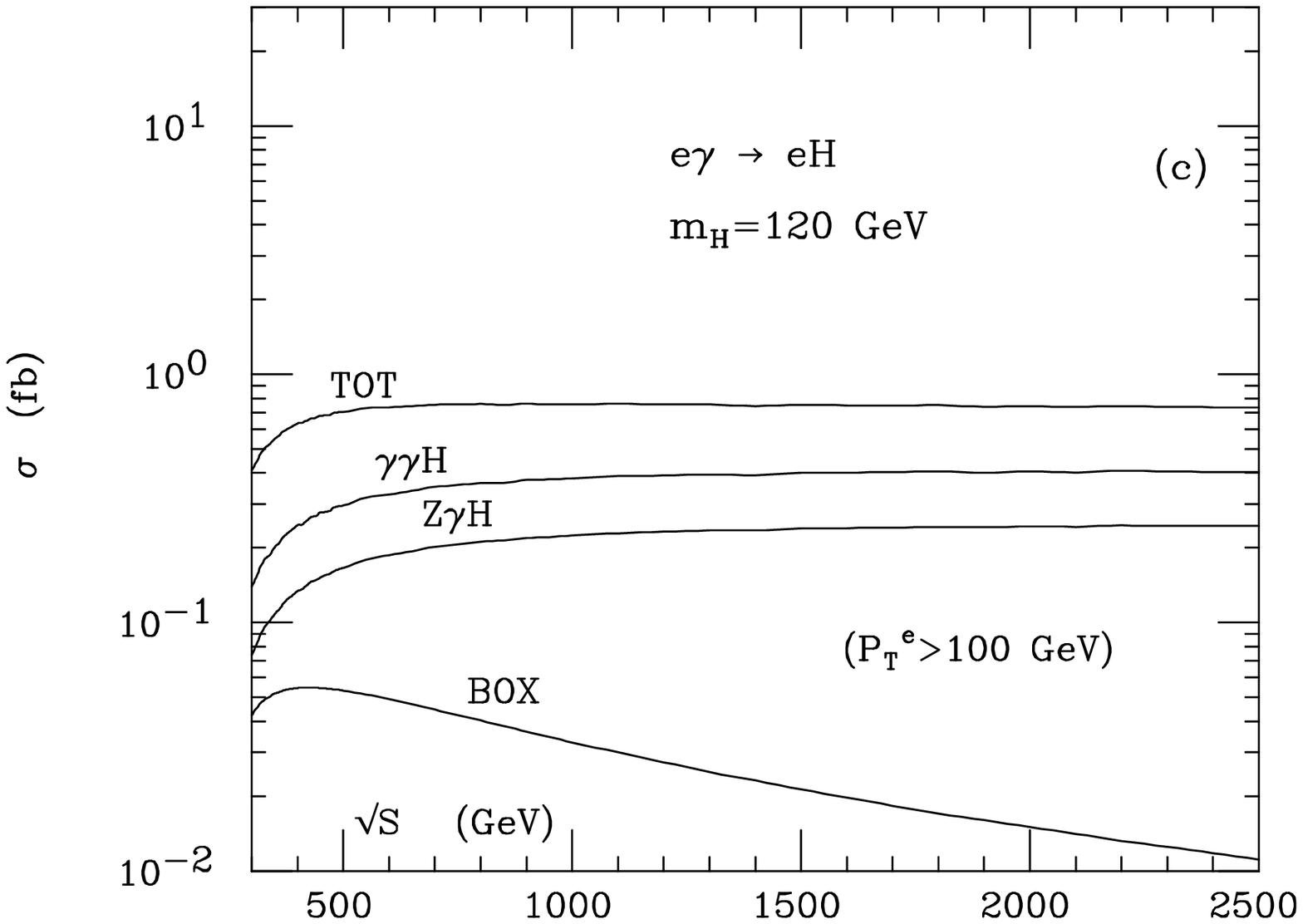}}
\vspace*{-4.cm}
\caption{ Effect of varying the $\pte$ cut on the \eaeh cross section.}
\label{fig34}
\end{center}
\end{figure*}

In figure \ref{fig34}, we show a possible strategy to enhance 
the \zah vertex effects in the $He$ production rate.
This consists in requiring a final electron (positron) tagged
at large angle. The corresponding cut on the transferred
squared momentum  $t$ depletes mainly the amplitudes involving a 
photon propagator in the $t$ channel. 
This can be easily seen from the three plots
in figure \ref{fig34}, where the cross sections dependence on $\sqs$,
for $\mh=120$GeV, is shown for no cut on the electron transverse
momentum $\pte$ (a), for a cut $\pte>10$GeV  (b),
and a cut  $\pte>100$GeV  (c).
The relative weight of the \zah and BOX contributions 
with respect to the total cross
section is considerably enhanced by a cut on the minimum allowed $\pte$.
For $\pte>100$GeV, \zah is about 60\% of \aah, and \zah gives
a considerable fraction of the total production rate, which still is
sufficient to guarantee investigation (about 0.7 fb).
One can also notice that the BOX contribution is of some relevance
only in the lower $\sqs$ range. 
\\
Note that the slight increase in the
\zah and BOX ``cross sections" when going from figure \ref{fig34} (a)
to figure \ref{fig34} (b) is due to the change of a factor
$\alpha(m_e)^2$ into $\alpha(\mw)^2$, which, as previously mentioned,
 we adopt for large-$\pte$ configurations.

We stress that, in the inclusive $He$ production, 
the bulk of the events are characterized by a forward final 
electron escaping detection. On the other hand, requiring 
a large $\pte$ corresponds, from an experimental point of view, 
to selecting a different final-state configuration, 
where the Higgs decay products have a large total transverse momentum,
balanced  by a high-energy  electron detected at large angle.

\noindent
\section{Background processes}
\vskip 10pt
Assuming  a final electron tagged at large transverse momentum in \eaeh,
 we now address the issue of separating
the signal coming from  an intermediate mass Higgs
(i.e., with 90GeV$\lta \mh \lta 140$GeV) 
from the most important  background channels. 
We recall that the main decay mode
for an intermediate mass Higgs is through the channel
$H\to b\bar b$, with a branching fraction of about 85\%.

An in-depth discussion of the problem has been presented in
\cite{ebol} in the different case of a collinear (undetected)
final electron, where one can adopt the WW approximation approach.
As we have already stressed, the latter is not useful
for distinguishing \zah vertex effects.

The main irreducible background to the process
\eaehb comes from the channel \eaebb. In the latter, a $b$ quark pair
is produced either through the decay of a virtual $\gamma(Z)$ 
or via the fusion of the initial $\gamma$ with a (virtual)
$\gamma$ or Z radiated by the electron beam.
The complete set of Feynman diagrams is given by
8 graphs and is shown in figure \ref{dgrm5}.

A crucial parameter to set the importance of the \eaebb background
is the experimental resolution on the $b\bar b$ invariant mass
$\delbb$. The background rates we present here are obtained
by integrating the $\mbb$ distribution over the range 
$\mh-\delbb<\mbb<\mh+\delbb$. We assume a very good mass resolution
on the $b$ quark pair, i.e. $\delbb=3$GeV.
Reaching a good resolution on $\mbb$ can be actually easier
in the $eH$ production at large angle.
Indeed, the tagging of the final electron $e_f$ implies the 
possibility of determining its energy with good accuracy. 
This reflects into an indirect (additional) determination of 
$\mbb$ through the relation
$$
E(e_{f})=\frac{s-\mbb^2}{2\sqs}.
$$
Assuming a monochromatic photon beam and neglecting 
ISR effects (i.e. assuming a fixed $s$), 
the latter implies a direct connection between
the $\mbb$ resolution and the $e_f$ energy resolution,
which can help in improving $\delbb$ in the 
final state configuration considered here (see also \cite{maet}).

We now carry out a detailed analysis of the
background from \eaebb. We use CompHEP to generate the
kinematical distributions and cross sections. As anticipated, all the 
rates presented are obtained by integrating the $\mbb$ distribution
over the range $\mh-\delbb<\mbb<\mh+\delbb$, with $\delbb=3$GeV.
As for the signal rates, we obtain the distributions for the process
\eaehb by  convoluting the $H$ distribution for \eaeh
with an isotropic (in the Higgs rest frame) 
decay $H\to b\bar b$, with proper branching ratio.
This chain, too, is  implemented in a modified version of CompHEP, that
generates events according to the exact one-loop matrix element for \eaeh.

\begin{figure*}[t]
\vspace*{-4.cm}
\begin{center}
 \mbox{\epsfxsize=16cm\epsfysize=18.5cm\epsffile{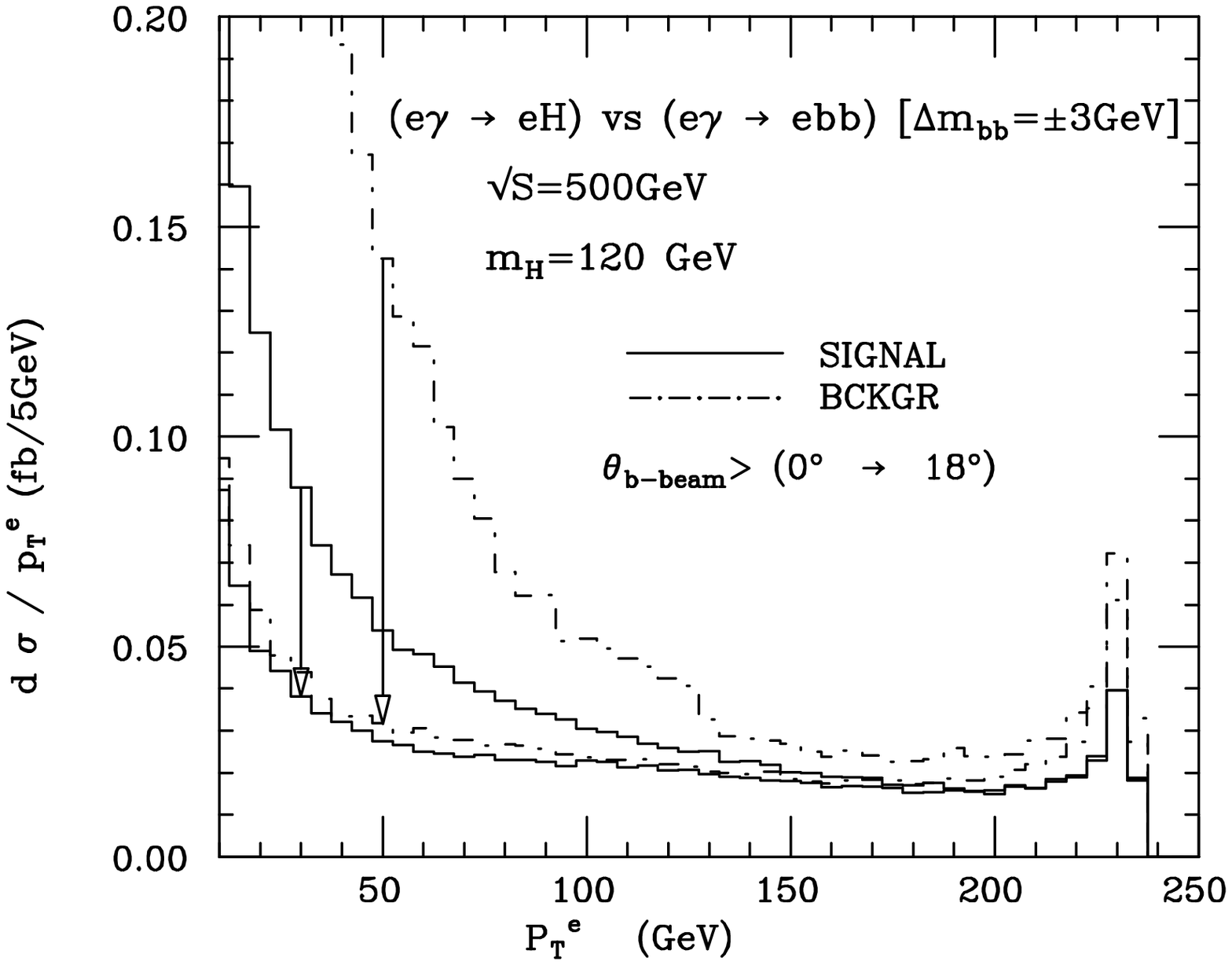}}
\vspace*{-5cm}
\caption{ Distribution in $\pte$ for the signal and the background
before and after angular cuts on the $b$'s is applied.
 }
\label{fig41}
\end{center}
\vspace*{-4.5 cm}
\begin{center}
 \mbox{\epsfxsize=16cm\epsfysize=18.5cm\epsffile{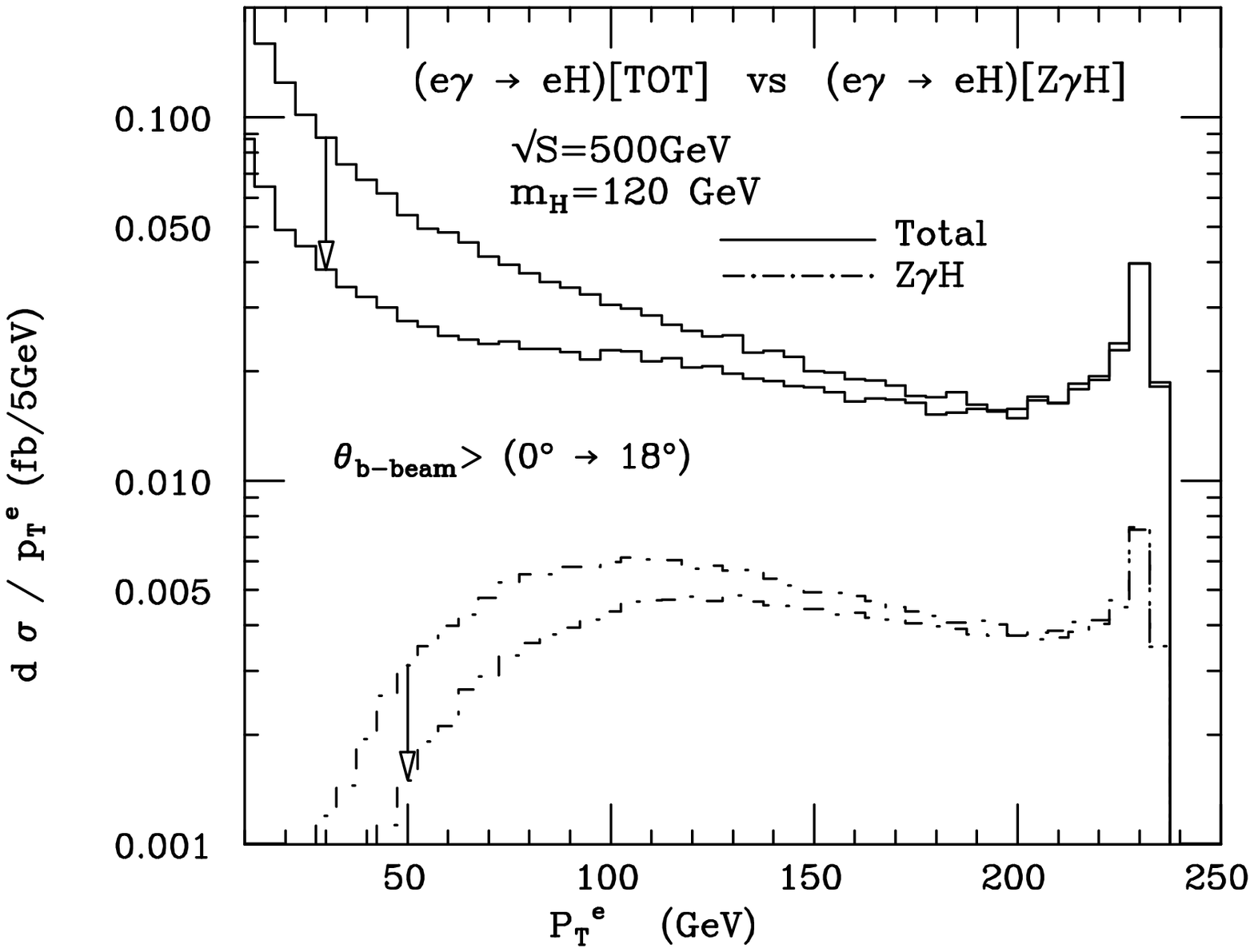}}
\vspace*{-4.9cm}
\caption{ Same as in the previous figure, for the signal and
the \zah contribution to the signal.
 }
\label{fig42}
\end{center}
\end{figure*}

In figure \ref{fig41}, the upper solid and dot-dashed histograms
show the $\pte$ distributions for the signal and background,
respectively, for $\mh=120$GeV and $\sqs=500$GeV. The background
is considerably larger than the signal, especially at 
moderate values of $\pte$. A possible way to improve this picture
is by putting a cut on the angles between each $b$ and the initial
beams. In fact, the vector couplings that characterize
the $b$'s in the channel \eaebb give rise to a $b$ angular
distribution considerably more forward-backward peaked than
in the case of the scalar $Hb\bar b$ coupling relevant for the signal.
In figure \ref{fig41}, the arrows show the lowering of the 
$\pte$ distributions, when an angular cut $\thebb>18^o$
is applied between each $b$ quark and both the beams.
This particular value of the angular cut reduces the signal and background 
distributions at a comparable level, without penalizing appreciably the
signal rate at large $\pte$. Note that the cut $\thebb>18^o$
has been optimized at $\sqs=500$GeV. Lower angular cuts will be
more convenient at larger $\sqs$.
 
Since, we are interested in isolating \zah effects,
in figure \ref{fig42} we compare the same  $\pte$ distribution of the signal 
(and the corresponding effect of the $\thebb$ cut) 
with the distribution coming from 
the pure squared \zah amplitude. One can check that the latter is 
concentrated at large $\pte$ values,
which is a typical effect
of the massive $Z$ propagator in the $t$ channel. 
The corresponding contribution to the total rate is practically unaltered 
if one imposes a cut $\pte\gta 50$GeV, 

A further source of background for the process \eaehb is the charm production
through \eaecc, when the $c$ quarks are misidentified
into $b$'s. This reducible background can be cured by a good $b$-tagging
efficiency, that should control a charm production rate that can be even more
than a factor 10 larger than the corresponding \eaebb cross section,
depending on the particular kinematical configuration
\cite{ebol}.
We computed the rate for \eaecc. By assuming a 10\% probability of
misidentifying a $c$ quark into a $b$  (hence, considering only a 
fraction 1/10 of the computed \eaecc rate), 
we find that this reducible background has lower rates
than the irreducible one. This can be seen in table \ref{tab41}, where
the signal is compared with both the reducible and irreducible background,
for two different sets of kinematical cuts, that enhance the \zah
contribution, and $\mh=$120 GeV, at $\sqs=$500 GeV.
Different initial polarizations for the $e$ beam are considered
(see section 5). For unpolarized beams and $\pte>100$GeV,
the \eaecc ``effective rate" is less than 1/3 on the \eaebb rate.
Note that the \eaecc channel is kinematically similar to
\eaebb. Hence, the particular strategies analyzed here to reduce the latter
authomatically depletes also the former.

A further background, that was considered in \cite{ebol}, is the resolved
\egebb production, 
where the photon interacts via its
gluonic content. Its estimate  depends on the particular assumption
for the gluon distribution in the photon, that is presently poorly known. 
We anyway tried to evaluate also this possible background,
by assuming that the gluon distribution in the photon beam
is given by the parameterization \cite{gluc} (where we have set
the energy scale $Q^2$ in the structure functions equal to 
$4 m_b^2$)\footnote{This choice of the scale among possible others
tends to maximize the resolved-photon rate.}.
For instance, for the same set of kinematical cuts, and the same $\mh$
and $\sqs$ values assumed in table \ref{tab41},
we have found that \egebb contributes to the background with
rates of 9.6 10$^{-3}$ fb and 0.40 fb, 
for $\pte>100$ and 10 GeV, respectively (and unpolarized beams). 
We also evaluated the contribution from
the $c$ quark production by resolved photons, by \egecc. Assuming,
as above, a 10\% probability of misidentifying a $c$ into a $b$,
and the same kinematical cuts, the corresponding rate are
2.8 10$^{-3}$ fb and 0.16 fb,
for $\pte>100$ and 10 GeV, respectively (and unpolarized beams).
The rates presented here derive from a leading-order
parameterization of the photon structure functions in \cite{gluc}. 
We checked
that a higher-order parameterization rises the results by at most 10\%.
Compared to the direct photon contributions reported in table \ref{tab41},
the resolved photon background should hence only marginally alter the
signal to background ratio, especially at large $\pte$
\footnote{The resolved photon rates above include only the gluon content of the 
initial photon beam.
The contribution coming from the gluon content 
of the virtual photon, that can be radiated by the initial electron, is not
included here. This is anyway expected to be less important than the former
contribution.}.

In the following, we will restrict to consider the irreducible
\eaebb background, being confident that,  at large values 
of the $\pte$ where the \zah effects are enhanced,
the latter provides the dominant component to the \eaeh background.

\noindent
\section{Beam-polarization effects}
\vskip 10pt
One of the advantages of a linear collider is the possibility
to work with polarized beams. This may allow, on the one hand, to test 
the parity structure of the interactions governing a particular
process and, on the other
hand, to optimize its background suppression.
Here, we consider the possibility of having either 
the electron or the photon beam longitudinally polarized.

\begin{figure*}[t]
\vspace*{-4.cm}
\begin{center}
\mbox{
\hspace*{-2.3cm}
 \mbox{\epsfxsize=11cm\epsfysize=18cm\epsffile{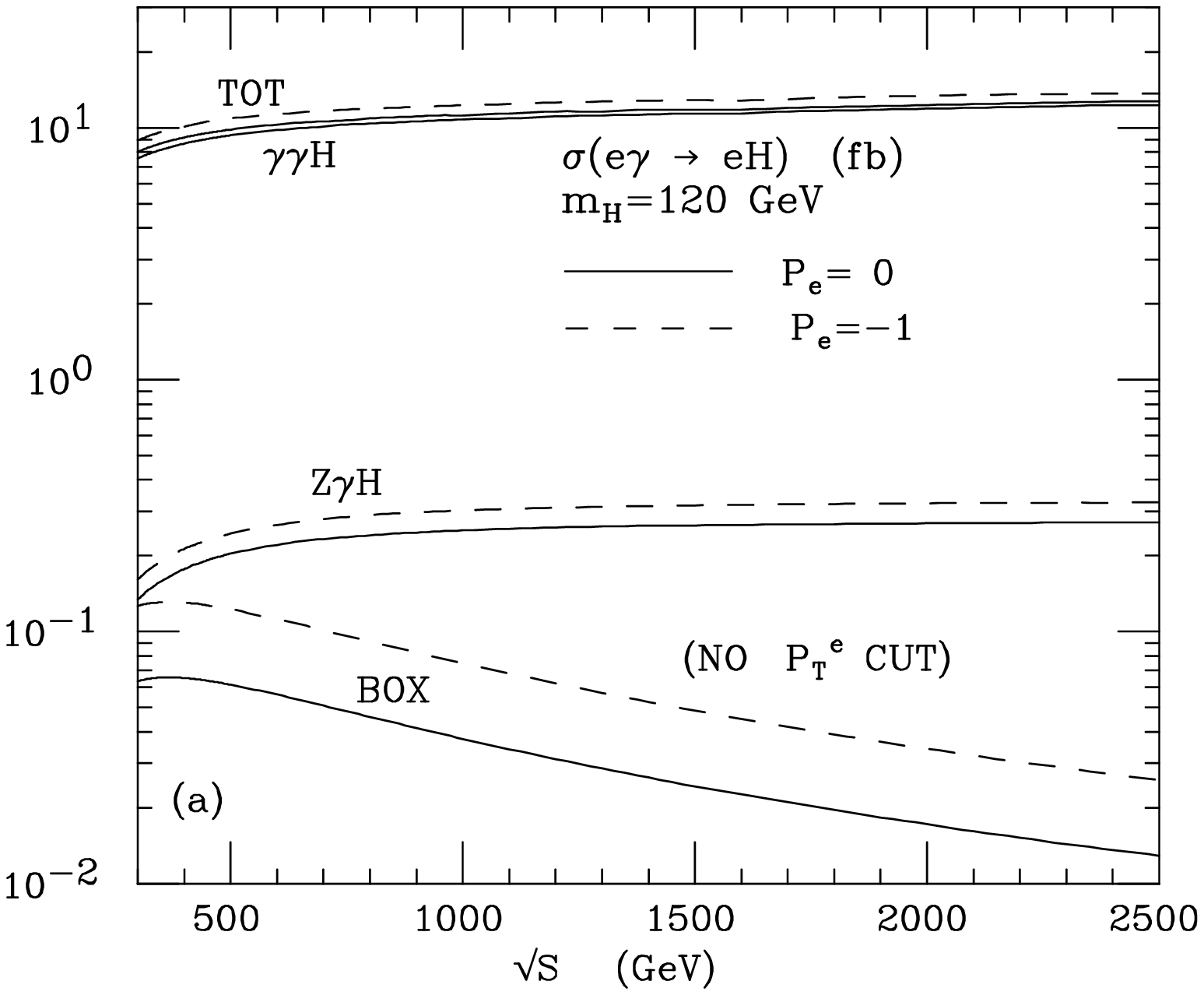}}
\hspace*{-2.2cm}
\mbox{\epsfxsize=11cm\epsfysize=18cm\epsffile{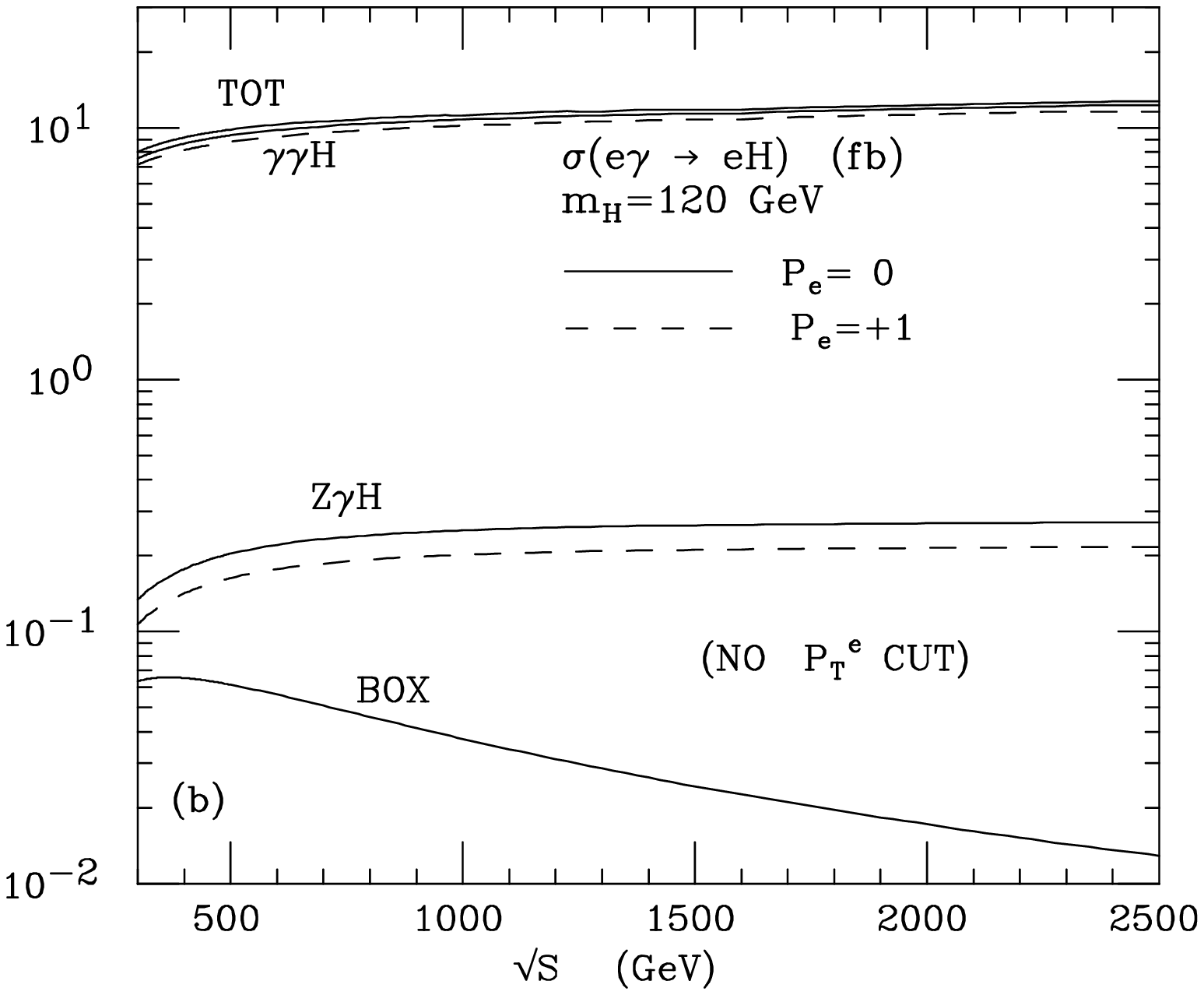}}}
\end{center}
\vspace*{-8.5cm}
\begin{center}
\mbox{
\hspace*{-2.3cm}
 \mbox{\epsfxsize=11cm\epsfysize=18cm\epsffile{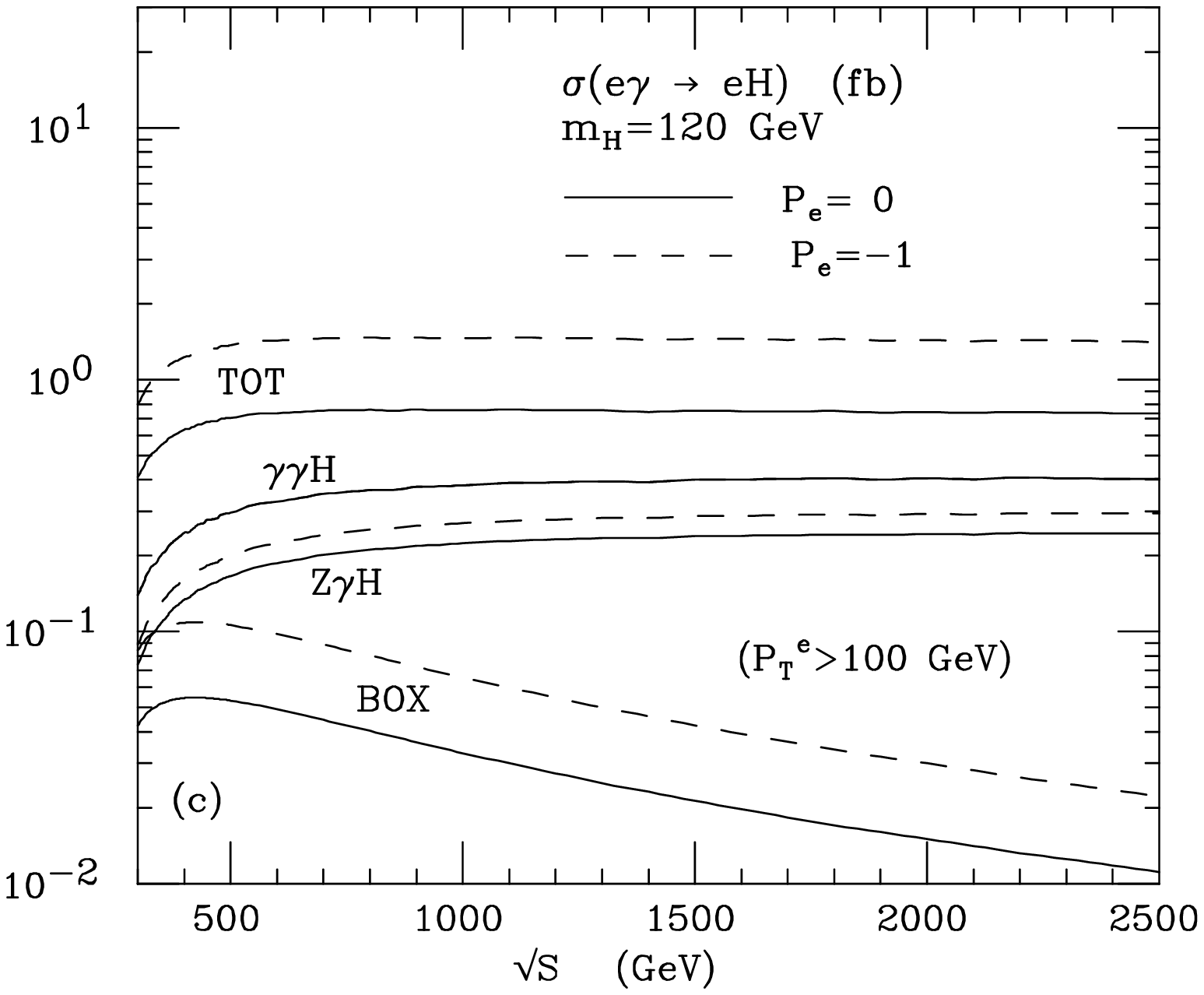}}
\hspace*{-2.2cm}
\mbox{\epsfxsize=11cm\epsfysize=18cm\epsffile{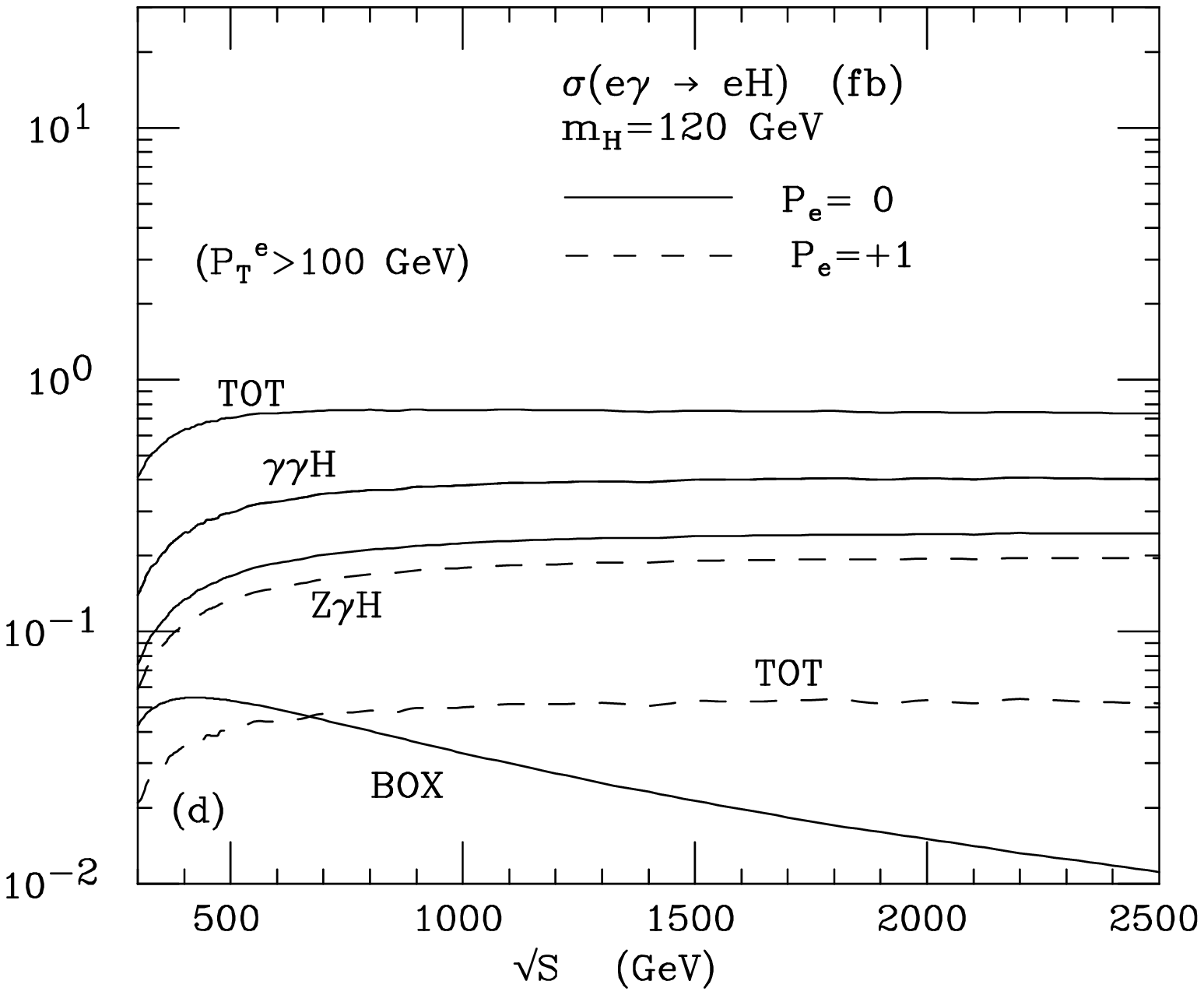}}}
\vspace*{-5.cm}
\caption{ Electron beam polarization effects  without
[(a) and (b)] and with [(c) and (d)] a cut $\pte>100$GeV.
 }
\label{fig51}
\end{center}
\end{figure*}

In figures \ref{fig51} (a) and (b), for $\mh=120$GeV and versus $\sqs$, 
we show the total cross section (and its \aah, \zah and BOX components) 
for the unpolarized case (solid) and
a completely longitudinally polarized electron (dashed).
In particular figure \ref{fig51} (a) refers to a left-handed 
electron beam ($\pe=-1$), while  figure \ref{fig51} (b) presents the case of
a right-handed electron ($\pe=+1$).
While the \aah curve is unaltered by a $\pe\neq 0$ value, the total
cross section is slightly modified by the influence of the electron
polarization on the parity non-conserving \zah and BOX couplings.
In particular, a left(right)-handed electron beam increases
(decreases) $\sit$, the \zah and the BOX contribution
by about 11\%, 20\% and 100\%, respectively, at $\sqs=500$GeV.
The strong variation in the BOX component is 
produced by the dominance of the $W$-box sector in this
contribution.

In figures \ref{fig51} (c) and (b), the same plots are given
when a cut $\pte>100$GeV is applied on the final electron
transverse momentum. One can see that in the high $\pte$ sector of the
phase-space, the total rates are much more sensitive to the electron
polarization.
For instance, assuming $\pe=-1$ ($\pe=+1$)  the total rate
increases (decreases) by about 94\% at $\sqs=500$GeV.

Some insight into this result can be gained by looking at tables 
\ref{tab53a} and \ref{tab53b},
where the $e/\gamma$ polarization dependence of the  interference 
pattern of the \aah, \zah and BOX contributions
is shown for $\pte>10$GeV and $\pte>100$GeV, respectively, at $\sqs=500$GeV.
For instance, one can see that for $\pe=+1$ there is a strong
destructive interference between the terms \aah and \zah.
This is essentially due to the different sign of the 
couplings $ee\gamma$ and $e_Re_RZ$, where $e_R$ stands 
for the right-handed electron component (see also section 2).

\begin{figure}[h]
\centerline{
\epsfig{figure=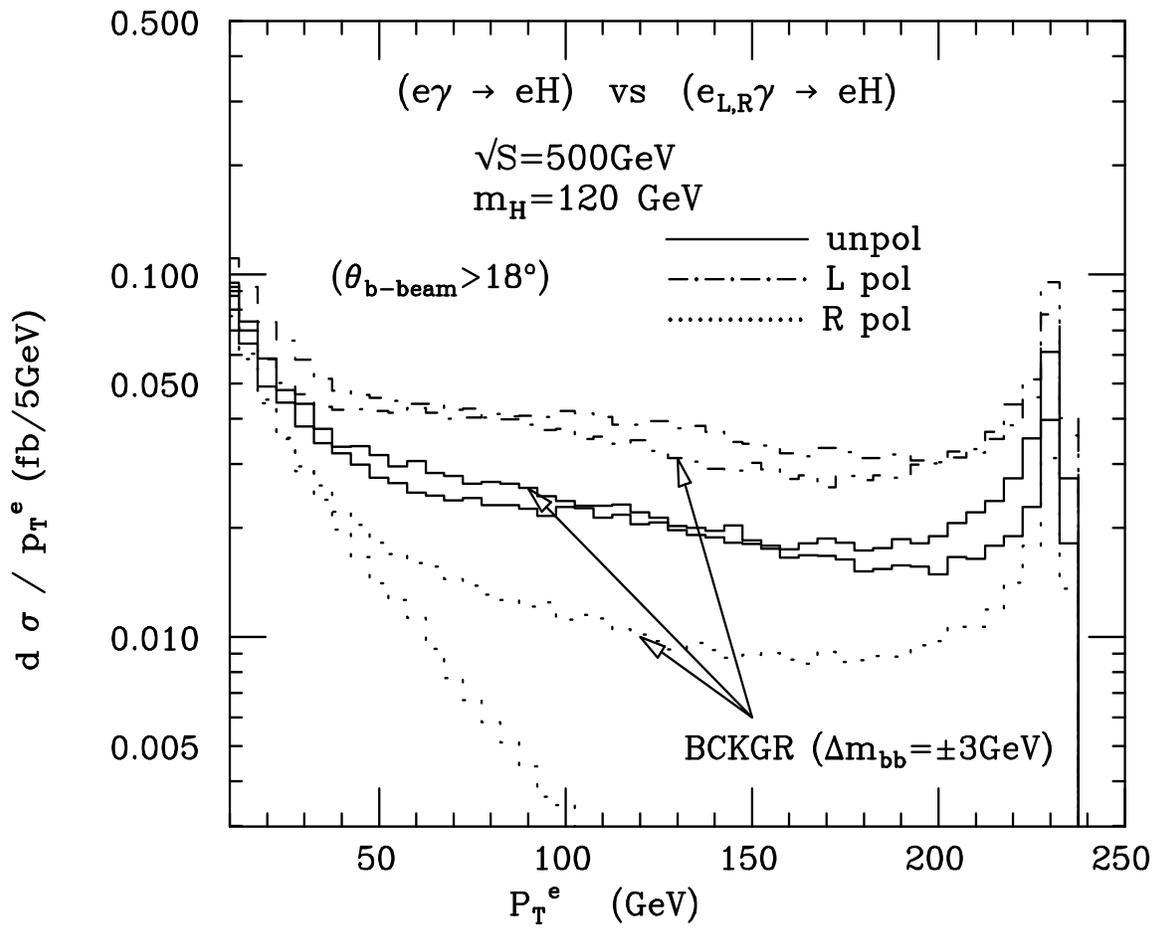,height=1.4\textwidth,
	width=\textwidth,angle=0}
}
\vspace*{-6.cm}
\caption{ Electron beam polarization effects on the
$\pte$ distributions.}
\label{fig52}
\end{figure}

The fact that a longitudinal polarization of the electron beam affects 
drastically the large $\pte$ range can also be clearly seen in 
figure \ref{fig52}, where the $\pte$ distributions relative to the
unpolarized and  to the left-handed and right-handed polarized
$e$ beam are presented for the signal and the \eaebb background.
One can also see that, although both the signal S and background B are
increased by a left-handed polarization, the ratio $S/B$ is improved at 
large $\pte$. 

\begin{figure*}[t]
\vspace*{-4.cm}
\begin{center}
\mbox{
\hspace*{-2.3cm}
 \mbox{\epsfxsize=11cm\epsfysize=18cm\epsffile{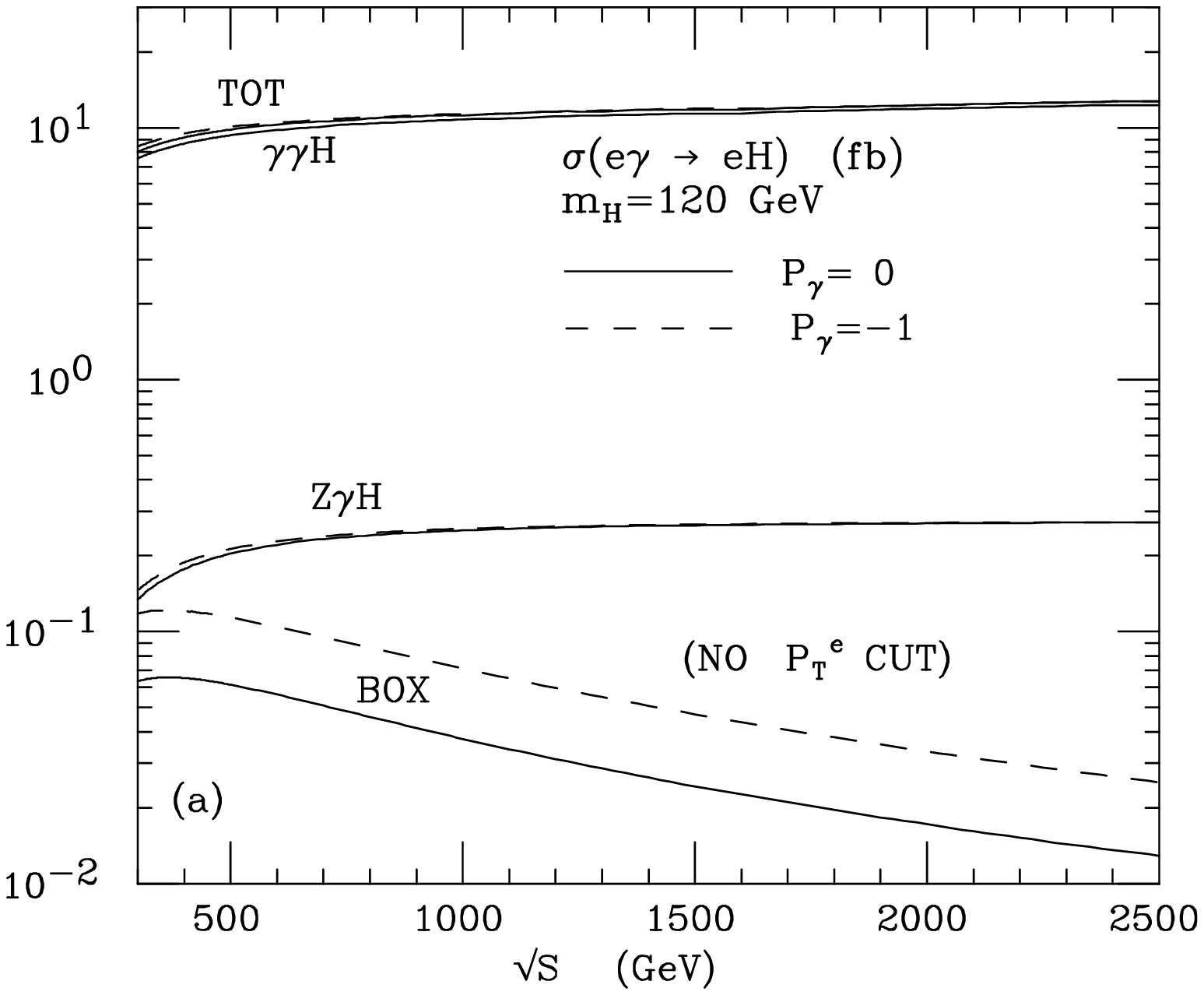}}
\hspace*{-2.2cm}
\mbox{\epsfxsize=11cm\epsfysize=18cm\epsffile{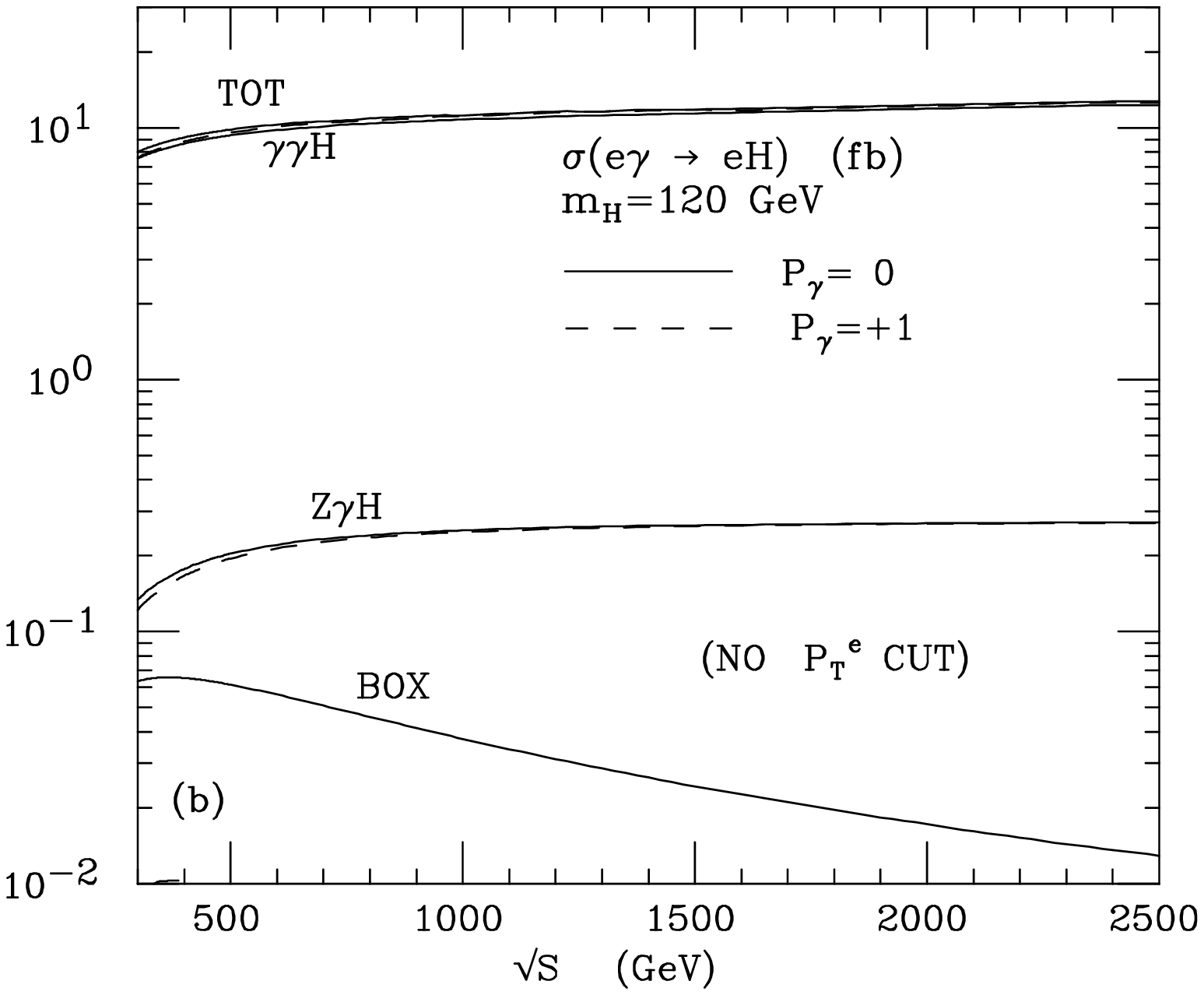}}}
\end{center}
\vspace*{-8.5cm}
\begin{center}
\mbox{
\hspace*{-2.3cm}
 \mbox{\epsfxsize=11cm\epsfysize=18cm\epsffile{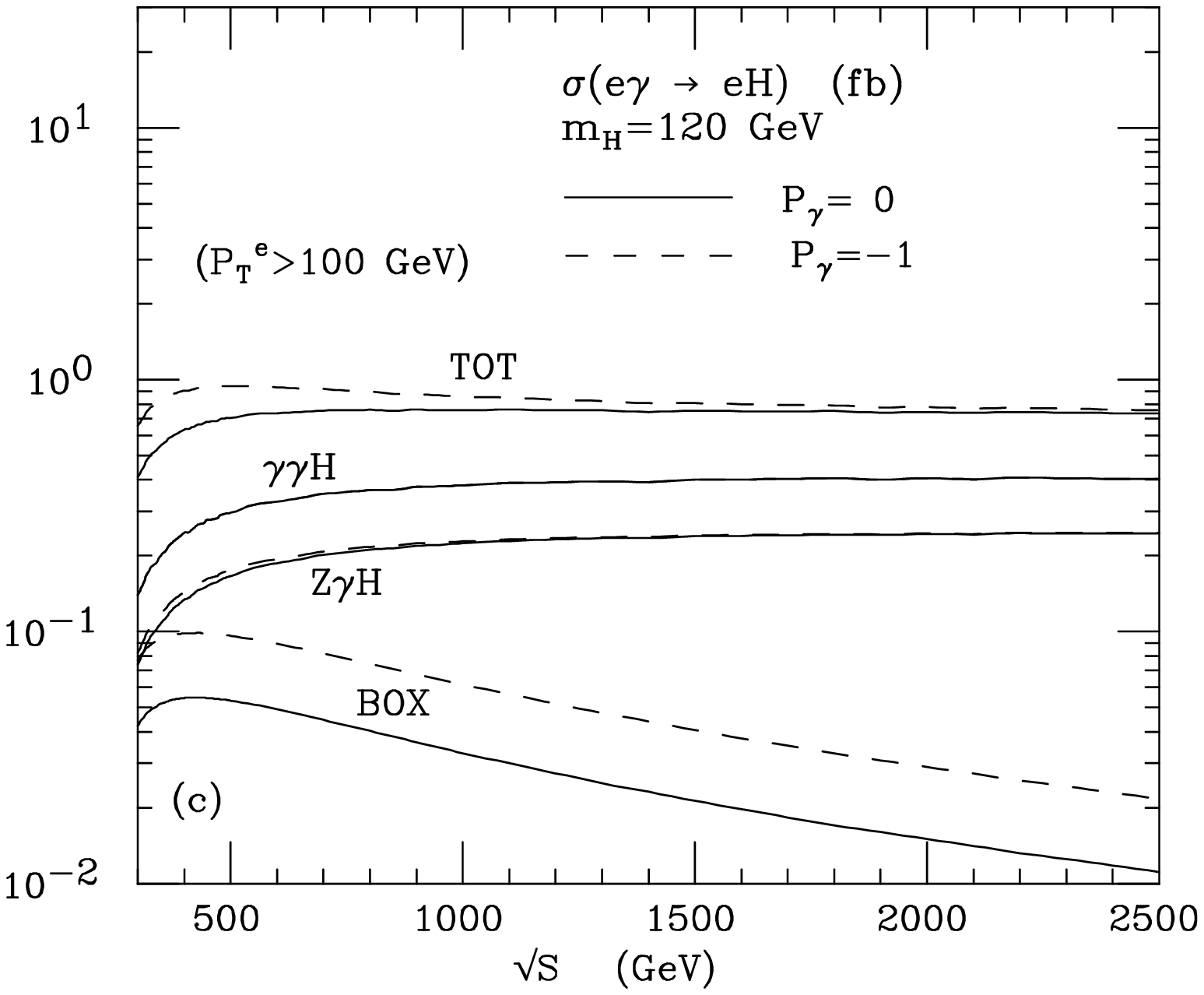}}
\hspace*{-2.2cm}
\mbox{\epsfxsize=11cm\epsfysize=18cm\epsffile{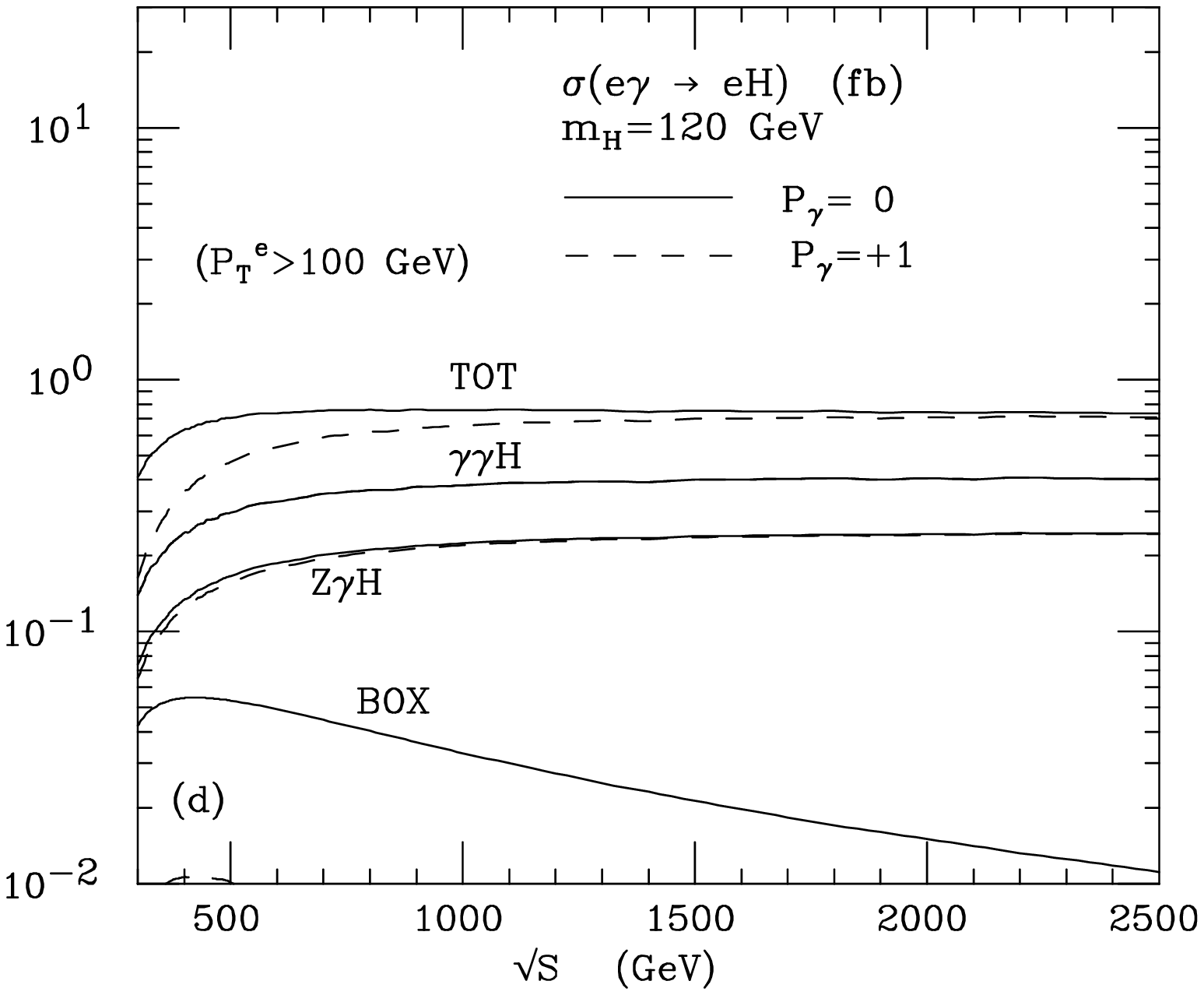}}}
\vspace*{-5.cm}
\caption{ Photon beam polarization effects  without
[(a) and (b)] and with [(c) and (d)] a cut $\pte>100$GeV.
 }
\label{fig54}
\end{center}
\end{figure*}

Figures \ref{fig54} (a), (b), (c) and (d) show the effects of assuming a
longitudinally polarized photon beam in the same framework of figures 
\ref{fig51}. The trend is similar to the polarized $e$ case, but the effect
is quantitatively more modest for a polarized $\gamma$, 
especially at large values of $\sqs$. 
The only exception is given by the BOX contribution
that is still considerably altered by $\pg\neq 0$ at any $\sqs$.
\noindent
\section{Initial State Radiation effects}
\vskip 10pt
The effects of the ISR on the signal and the background rates can be 
taken into account by  folding  the corresponding 
cross sections with a structure function describing the reduction
of the electron beam energy because of the QED radiation.
We adopt the approach of \cite{isrr}, that 
is accurate at the next-to-leading order for collinear emission 
and resums soft photon effects.
All this is implemented through the computer package CompHEP, which
automatically takes into account also the kinematical cuts needed
either to enhance the \zah contribution in the signal (i.e., $\pte$ cuts)
or to decrease the relative importance of the background
(i.e., $\thebb$ cuts).

We compared the rates of the 
unfolded cross sections  with the cross sections convoluted 
with the ISR structure function. The effect of the ISR 
in our context has been found to be marginal in general. In particular, 
we found  that the ISR effects slightly reduce the 
signal for all the electron polarization states. For the kinematical
configurations described in table \ref{tab41}, the signal is reduced by 
about 3\%, for $\pte>100$GeV and even less (about 1\%) for $\pte>10$GeV.
On the other hand, the irreducible \eaebb
background is a little enhanced by the ISR. For 
$\pte>100$GeV, it is increased by less than 1\%, while, for $\pte>10$GeV,
it raises by about 4\%.
This holds for both $\pe=0$ and $\pe=\pm 1$.

We checked that such behaviors can be easily explained in terms of the 
increasing/decreasing of the relevant cross section with $\sqs$,
when all the relevant kinematical cuts are taken into account.

Altogether, one can conclude that the ``unfolded" general picture is
only mildly modified by the ISR effects.
 
\noindent
\section{Optimization of S/B and electron asymmetries}
\vskip 10pt

As can be seen in table \ref{tab41}, a $\pte$ cut
of 100 GeV, along with a resolution on $\mbb$ of $\pm 3$ GeV and
a cut on all the $b$'s that are closer than 18$^o$ to the beams,
optimizes the $S/B$ ratio,
for $\mh=$120 GeV and $\sqs=$ 500 GeV. In particular, it
gives rise to a signal rate of 0.40 (0.78) fb versus
an irreducible background rate of 0.63 (0.96) fb in the unpolarized 
($\pe=-1$ polarized) case.
This means that the signal and the background are comparable
in the interesting configurations. With an integrated luminosity of
100 pb$^{-1}$, the corresponding statistical significance of
the signal  from \eaeh is of the order $S/\sqrt{S+B}\simeq 4$ for 
$\pe=0$ and 6, if $\pe=-1$.
This implies the possibility of measuring the 
corresponding cross section for \eaeh 
with an accuracy of about 25\% (17\% for polarized $\pe=-1$ beam), 
unless systematic errors dominate.

Of course, 
the final accuracy on the determination of the coupling of the 
Higgs boson to the $Z$ and $\gamma$  is not simply the accuracy on 
the cross-section measurement.
As we can see from  table \ref{tab53b},  for $\pte>100$GeV and $\sqs=500$GeV 
(and  unpolarized beams),
the  \zah  vertex contributes about half of the measured cross sections,
including the interference effects.
Assuming that the Higgs boson coupling with the photons is tested 
and  measured with high accuracy in some different
process, the statistical sensitivity to the Higgs  coupling with the $Z$
and $\gamma$  gets of the order $\frac{1}{2} S/\sqrt{S+B}$
(note that, assuming a reduced integrated luminosity of 
$r100$ pb$^{-1}$, with $r<1$, would in general lower the expected accuracy
by a factor $\sqrt{r}$).

\begin{figure}[h]
\centerline{
\epsfig{figure=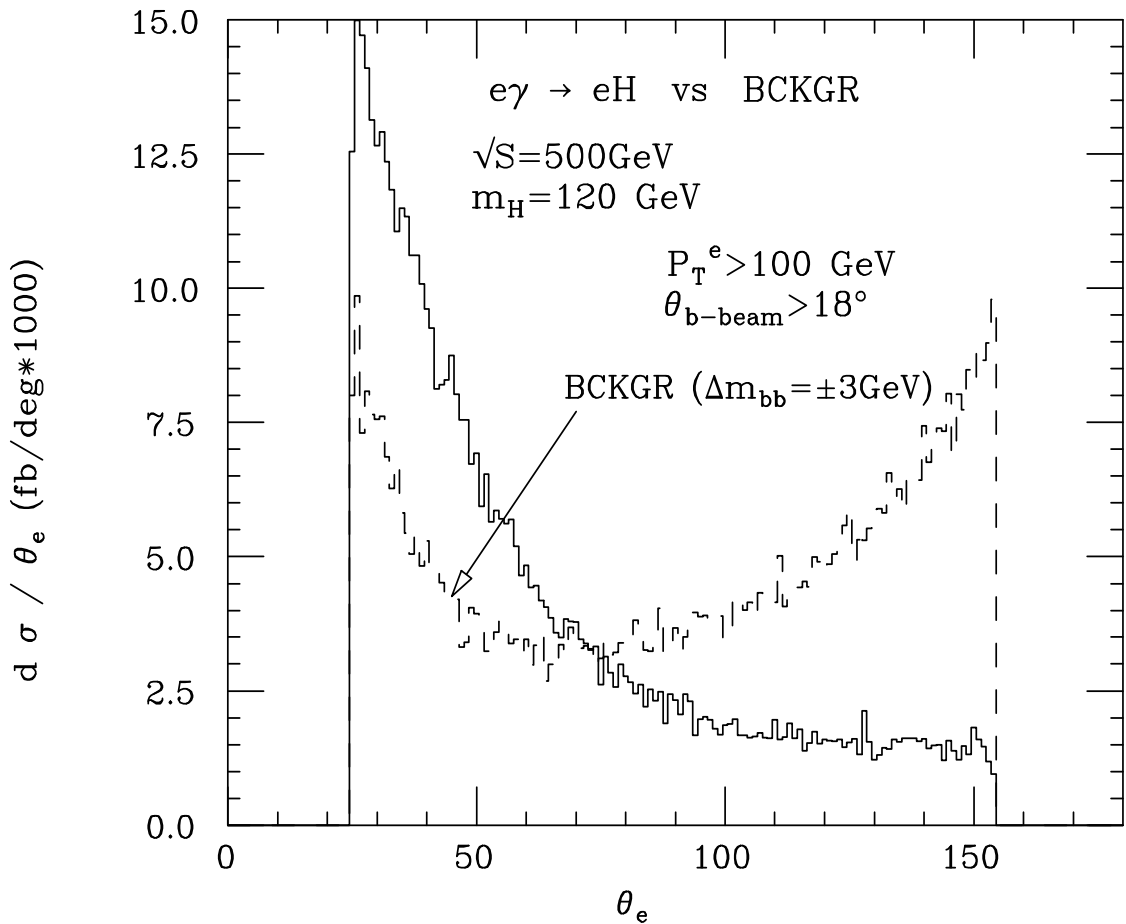,height=1.4\textwidth,
	width=\textwidth,angle=0}
}
\vspace*{-6.cm}
\caption{Final electron angular distribution with respect 
to the initial electron beam.
The solid (dashed) line refers to  the signal
(irreducible \eaebb background).
The kinematical cuts applied are shown in the plot.
The initial beams are assumed to be unpolarized.}
\label{fig71}
\end{figure}

There is a further way to improve the accuracy 
on the cross section measurement.
This is  by exploiting the
electron angular asymmetry of the signal  with respect to the beam. 
Indeed, we found
that in the \eaebb background the final electron angular distribution,
although not completely symmetric, is almost equally shared in the forward and
backward direction with respect to the beam.
In particular, we checked that the 2nd diagram in 
figure 5 is responsible for the backward peak, while the 4th diagram 
gives the forward one.
\\ 
On the contrary, the final
electron in \eaeh is mostly directed in the forward direction.
The  typical behavior  is shown in figure \ref{fig71}, where
the solid (dashed) line gives the histogram for the final electron
angular distribution [in the centre-of-mass system] 
for the signal (background),
for $\pte>100$GeV and  $\theta(b-beam)>18^o$,
at $\mh=120$ GeV and $\sqs=500$GeV.
As usual, the background is integrated over the range 
$\mh-\delbb<\mbb<\mh+\delbb$ with $\delbb=3$GeV. 
The initial beams are assumed to be unpolarized. 
The strong asymmetry in the signal $\theta_e$ distribution
is manifest. The final electron is mostly scattered
forwards in \eaeh (the empty intervals for $\theta_e\lta 25^\circ$
and $\theta_e\gta 155^\circ$ 
are just due to the $\pte>100$GeV cut). 
Note that this pattern keeps valid also 
when relaxing the $\pte$ cut, and for polarized beams.

It is straightforward at this point to enhance the $S/B$ ratio, by
simply measuring the difference between the forward and backward 
cross sections. In table \ref{tab71}, after applying the same cuts as in 
table \ref{tab41} and for different polarizations of the $e$ beam, we report
apart from the total rate, the forward cross sections 
and the difference of the forward and backward cross sections
${\cal S}_{FB}=\sigma(\theta_e<90^\circ)-\sigma(\theta_e>90^\circ)$.
One can see that, in the difference ${\cal S}_{FB}$,  80\% of the signal
survives, while the background is reduced by about an order of magnitude,
in both the interesting  $\pe=0$ and $\pe=-1$ cases.
With a luminosity of 100 pb$^{-1}$, one then gets an accuracy on
${\cal S}_{FB}$ of about 16\% for unpolarized  $e$ beams, and
12\% for  $\pe=-1$ (corresponding to 
$S/\sqrt{S+B}\simeq 6.4$ and 8.5, respectively.)
By the way, it could also be convenient to measure the
relative asymmetry  ${\cal S}_{FB}/
[\sigma(\theta_e<90^\circ)+\sigma(\theta_e>90^\circ)]$, 
that has the advantage of being free from
possible uncertainties on the absolute normalization of the cross sections.

The analysis above can be also extended  to the  study of
possible  anomalous contributions in the \zah amplitude 
coming from some  extension of the standard model.
For instance, all models causing a variation in the cross section and/or
angular asymmetries by more than about 20\% should be easily disentangled 
in the same experimental conditions analyzed here.
Furthermore, since in the latter case what we call here signal could act
as a further background, it could be convenient to 
consider also the $\pe=+1$ cross sections \cite{nuov}. 
Indeed, as shown in table \ref{tab53b},
for large $\pte$ the right-handed polarized $e$ beam
minimizes the standard model \eaeh {\it background}, because of 
the strong negative interferences between the different amplitudes.

\noindent
\section{Conclusions}
\vskip 10pt
The study of the exact rates for the process \eaehb in the intermediate
$\mh$ range confirms that the associated $He$ production in 
$e\gamma$ collisions is a competitive means with respect to the process
\aatoh to study the vertex \aah and its possible anomalies
(as anticipated by the analysis made in the WW approximation \cite{ebol}).
The relevant total (unpolarized) cross sections 
are in the range $(9\div 17)$fb,
for $\mh=(90\div 150)$GeV and $\sqs=(0.5\div 1.5)$GeV, which, 
assuming an integrated luminosity of 100 fb$^{-1}$, corresponds
to ${\cal O}(10^3)$ Higgs events.

If the final electron is tagged at large $\pte$, 
a further possibility offered by the channel \eaeh  
is to study the effects coming from the \zah vertex,
keeping still a reasonable statistics [${\cal O}(10^2)$ events]. 
This possibility 
has not any counterpart in the \epem and \gag collision physics.
Graphs with boxes, too, contribute at large $\pte$, but their 
relative importance decreases with $\sqs$ for $\sqs>400$GeV.

We checked that the main background comes from the process \eaebb. This can 
be controlled  by a good $\mbb$ experimental resolution
(that can be improved by the final electron energy determination),
and by requiring that the final $b$ quarks be not too close to 
the beams direction.

We have also shown that starting with a left-handed polarized 
electron beam  doubles the rates and improves the $S/B$ ratio,
in the $\pte\gta 100$GeV kinematical range interesting for the \zah 
vertex studies. Further improvements in the $S/B$ ratio can be obtained by
exploiting the final-electron angular asymmetry of the signal.

The inclusion of the initial state radiation effects
marginally deteriorates the $S/B$ ratio.

The value of the integrated luminosity assumed in this study 
(that is 100 fb$^{-1}$)
does not seem to be essential to disentangle a \zah effect, although
a high luminosity 
would be crucial to increase the accuracy of the measurement.
With a luminosity of 100 fb$^{-1}$, one expects an accuracy as good as
about 10\% on the measurement of the \zah effects, at $\sqs=500$GeV.
A luminosity of  50 fb$^{-1}$ would anyhow allow to
measure the standard model signal with an accuracy better than 20\%.

In conclusion, the \eaehb turns out to be an excellent means
to check the standard-model 
one-loop coupling \zah. Further investigation
and comparison with the predictions from  possible extensions of the 
standard model altering the \zah vertex is worthwhile. 
This, we are planning to do in a
forthcoming paper \cite{nuov}.

\vspace{0.7cm} 
\noindent 
{\large \bf Acknowledgements} 

\noindent
Useful discussions with I.F.~Ginzburg and A.E.~Pukhov  are gratefully 
acknowledged.
We also wish to thank I.P.~Ivanov and A.T.~Banin for their independent check
of our results.
The work of V.A.I. was partly supported by the Grants INTAS-93-1180ext and
96-02-18635 from the Russian Fund for Basic Research. V.A.I. wishes also to
thank INFN for its hospitality and the possibility to work 
during his visit at the University of Roma 1.
\vskip 24pt
\newpage
\noindent
{\Large \bf Appendix} 

In this appendix, we give the explicit expressions for
the functions $B_0$, $C_0$ and $D_i$ appearing in
eqs. (\ref{tfunc}), (\ref{fwbox}) and (\ref{fzbox}).
Next, we give a short discussion on some sources of numerical 
instabilities arising in the computation of the total cross section, and 
describe the method used to control them.

\begin{itemize}
\item {\bf Loop integrals}
\end{itemize}

The complete set of independent loop integrals used 
in our calculations is defined by the following formulas 
in  dimensional regularization\footnote{Relations to other 
definitions of one-loop integrals:
\begin{itemize}
 \item Our definition of integrals corresponds to the {\it Passarino-Veltman} 
integrals \cite{passvelt} 
if metric signature is changed from $(-+++)$, used by Passarino-Veltman,
to $(+---)$. They also used a different choice of external momenta, 
hence their decomposition (\ref{decomp}) differ from ours.
\item We differ from \cite{dend} by an opposite sign of the $C$-integrals.
\end{itemize}
}, where the time-space dimension is $n=4-2\varepsilon$:
\bea
 B_0(p^2,m^2,m^2) &\equiv & \frac{1}{i\pi^2} \int \frac{d^n q} 
                   {[m^2-q^2]\, [m^2-(q+p)^2]}\; , \\
 C_0(m,p_1,m,p_2,m) &\equiv & 
    \frac{1}{i\pi^2} \int \frac{d^n q}
    {[m^2-q^2]\, [m^2-(q+p_1)^2]\, [m^2-(q+p_2)^2]}\; , \\
 D_{0;\mu;\mu\nu}(m_0,p_1,m_1,\ldots p_3,m_3) &\equiv & 
   \frac{1}{i\pi^2} \int \frac{\{1;q_\mu;q_\mu q_\nu\}\;d^n q}
  {[m_0^2-q^2]\, [m_1^2-(q+p_1)^2]\ldots[m_3^2-(q+p_3)^2]}.
\eea
Of course, each denominator factor originated from the propagators should be
treated in the Feynman limit:
$(M^2-p^2)^{-1} \equiv \lim\limits_{\epsilon \to 0}
    (M^2-p^2-i\epsilon)^{-1}$. 
As a result, these loop integrals have complex values in general.

We decompose the vector and tensor box-integrals over the covariant Lorentz 
structures:
\bea
D^\mu        &\equiv &  
      p_1^\mu\cdot D_1 \,+\,p_2^\mu\cdot D_2 \,+\,p_3^\mu\cdot D_3\, ,
             \nonumber \\
D^{\mu\nu}  &\equiv &  
      g^{\mu\nu}\cdot D_{00} \,+\, p_1^\mu p_1^\nu\cdot D_{11}
      \,+\,p_2^\mu p_2^\nu\cdot D_{22}\,+\,p_3^\mu p_3^\nu\cdot D_{33}  
                                                 \label{decomp}\\
   &  & +\,(p_1^\mu p_2^\nu+p_2^\mu p_1^\nu)\cdot D_{12}
       \,+\,(p_1^\mu p_3^\nu+p_3^\mu p_1^\nu)\cdot D_{13}
       \,+\,(p_2^\mu p_3^\nu+p_3^\mu p_2^\nu)\cdot D_{23}\, .  \nonumber
\eea

Note that all these integrals are UV finite. The only exception is
$B_0$, for which $\lim\limits_{\varepsilon\to 0} \varepsilon B_0=1$.
In our results, the integrals $B_0$ appear in the triangle contributions
only in a UV finite combination [see eq.~(\ref{tfunc})].

For the numerical evaluation of the scalar loop integrals, 
as well as of the scalar factors 
in the decomposition (\ref{decomp}), we used the FF library 
\cite{FF}, and the corresponding Fortran routines. Note that in our definition
of the loop integrals, the $C$-type integrals have an opposite sign
with respect to the latter. Moreover, in the
decomposition (\ref{decomp}) we use a set of external momenta different
from the FF library. 
Hence, we perform a linear transformation of our
scalar factors $D_i$ and $D_{ij}$, in order to get a connection
with the corresponding factors defined in the FF library.

Regarding the analytical evaluation of the amplitudes, 
the  check of the QED gauge--invariance 
identities (\ref{qedidentity}), as well as the extraction of the
QED gauge non-invariant terms (\ref{nig1}) and the corresponding 
ones coming from the $W$-box diagrams,
we used the computer algebra system REDUCE \cite{REDUCE}. 
 
\begin{itemize}
\item {\bf Numerical instabilities}
\end{itemize}

Although, in general, we computed the relevant amplitudes
in the $m_e=0$ chiral limit,
in the calculation of the total cross 
section we assumed for $t_{max}$ the exact value, that is approximately 
equal to $(- m_e^2\mh^4/s^2)$.
Then, when integrating the region near the $t$-channel pole at $t\simeq 0$
with routines in double precision for arithmetic operations,
we met some numerical instabilities.
In particular, we observed a lost of numerical precision 
in the evaluation of the kinematics for $t\ge -10^{-9} \mbox{GeV}^2$. 
In order to avoid this instability, we used the following procedure.
We introduced some parameter $t_0$ and approximated the matrix element $M$
for the $t$-channel photon contribution by $M=X(t)/t$ for $t\le t_0$, 
and $M=X(t_0)/t$ for $t_0\le t\le t_{max}$, where $X(t)$ is the numerator 
of the matrix element. 
If the parameter $t_0$ is taken close to $t_{max}$,
the latter turns out to be a good 
approximation,  since then $X(t_{max})$ differs very little 
from $X(t_0)$. To test this method, we checked the 
independence of the total cross section from the parameter $t_0$,
 by varying $t_0$ 
in the interval $-1\,\mbox{GeV}^2 \leq t_0 \leq -10^{-6} \mbox{GeV}^2$.

There is also a second source of numerical instability when the $W$ and 
$Z$-box diagrams are evaluated near $t=0$ and $u=0$.
When the tensor integrals for the boxes are expressed via the scalar integrals
by Passarino-Veltman \cite{passvelt}
(e.g., FF library uses this procedure \cite{FF}) some
spurious poles can arise from different terms. Of course, in the total 
results these poles cancel each other. In our case such spurious poles arise
at $t=0$ and $u=0$. We checked analytically the corresponding cancellation
in our results for the boxes (\ref{box},\ref{wzbox}). First, we
expressed the formulas (\ref{wzbox}) via scalar integrals
and then analyzed the final expressions in $t=0$ and  $u=0$, 
where they must vanish.
This analysis was made with 
the help of the REDUCE program PV \cite{PV}, that implements
the Passarino-Veltman procedure. 
Anyhow, it is not possible to check explicitly the cancellation of
these spurious poles at the level of the scalar integrals.
As a result, these integrals cannot be evaluated with 
good accuracy at $t\ge -10^{-6}$GeV$^2$,  since  
the numerical precision of the FF library is not sufficient to get this 
cancellation. 
Fortunately,  one can neglect the box contributions at such small $t$ and 
$u$, since these functions have a regular behaviour near $t=0$ and $u=0$.
Therefore,  we just  neglected the $W$ and $Z$-box contributions for 
$t,u\ge -10^{-2}$GeV$^2$.

\newpage

%
\begin{table}[htbp]\centering
\vspace*{1cm}
\begin{center}
\begin{tabular}{|c||c|c|c|c|c|c|}
\hline
 $m_H$
&\multicolumn{2}{c|}{$\sqrt s =$0.5 TeV} 
&\multicolumn{2}{c|}{$\sqrt s =$1 TeV} 
&\multicolumn{2}{c|}{$\sqrt s =$1.5 TeV} \\
\cline{2-7}
 (GeV) 
& $e\gamma\to e H$& $e\gamma\to \nu W H$ & 
$e\gamma\to e H$& $e\gamma\to \nu W H$ & 
$e\gamma\to e H$& $e\gamma\to \nu W H$ 
\\
\hline
  80    &     8.38     & 47.3   & 9.29   & 157.   &  9.74  & 243.   \\
  100   &     8.85     & 41.3   & 9.94   & 148.   &  10.5  & 233.   \\
  120   &     9.80     & 35.4   & 11.2   & 138.   &  11.8  & 223.   \\
  140   &     11.8     & 29.8   & 13.7   & 129.   &  14.6  & 212.   \\
  160   &     21.1     & 24.6   & 25.0   & 120.   &  26.6  & 201.   \\
  180   &     20.9     & 19.9   & 25.3   & 111.   &  27.0  & 191.   \\
  200   &     17.2     & 15.7   & 21.2   & 102.   &  22.8  & 181.   \\
  300   &     5.97     & 2.87   & 8.53   & 65.4   &  9.43  & 136.   \\
  400   &     1.64     & 0.0151 & 2.78   & 38.8   &  3.18  & 100.   \\
  500   &              &        & 0.501  & 20.9   & 0.595  & 72.9   \\
  600   &              &        & 0.0767 & 9.51   &0.0952  & 51.5   \\
  700   &              &        & 0.0608 & 3.23   &0.0901  & 35.0   \\
\hline
\end{tabular}

\end{center}
\caption{{\em Total cross sections in fb.
}}
\label{tab32}
\vspace*{-.3cm}
\end{table}

\begin{table}[b]
\begin{center}
\begin{tabular}{|c||c|c|c||c|c|c|}
\hline
$m_H=120$~GeV
&\multicolumn{3}{c||}{$p^e_T > 100\,$GeV}
&\multicolumn{3}{c|}{$p^e_T > 10\,$GeV}\\
\cline{2-7}
$\sqrt s = 500$~GeV 
& $\sigma(e H)\,(fb)$ & $\sigma(e b\bar b)\,(fb)$ & $\sigma(e c\bar c)\,(fb)$ 
& $\sigma(e H)\,(fb)$ & $\sigma(e b\bar b)\,(fb)$ & $\sigma(e c\bar c)\,(fb)$ 
\\
\hline
$P_e = 0$  & 0.404  & 0.634 & 0.208 & 1.01  & 1.34 & 0.868  \\
\hline
$P_e = -1$ & 0.780   & 0.961 & 0.277 & 1.60  & 1.94 & 1.00  \\
\hline
$P_e = +1$ & 0.0263 & 0.304 & 0.136 & 0.429 & 0.767 & 0.726 \\
\hline
\end{tabular}
\end{center}
\caption{{\em  Comparison of the signal with the irreducible
background $e \gamma \to e b\bar b$ and the reducible background
coming from $e \gamma \to e c\bar c$, for different $e$-beam polarizations. 
For the $e \gamma \to e c\bar c$ background a 10\% probability of 
misidentifying a $c$ quark into a $b$ is assumed
(that is, only 1/10 of the cross section is reported).
Two configurations for kinematical cuts are considered.
The angular cut $\theta(b(c)-beam)>18^o$ is  applied everywhere.
The signal rates includes 
the complete treatment of the $H\to b \bar b$ decay.
The $b\bar b$ invariant mass
for the background is integrated over the range 
$\mh-\delbb<\mbb(\mcc)<\mh+\delbb$ with $\delbb=3$GeV. 
}}
\label{tab41}
\end{table}

\begin{table}[b]
\vspace*{1cm}
\begin{center}
\hspace*{-1.5cm}
\begin{tabular}{|c|c||c|c|c|c|c|c|c|}
\hline
$\sqrt s =$& $m_H$
&\multicolumn{7}{c|}
{$\sigma(e\gamma\to e H, \;\; p^e_T > 10\,$GeV)$\;\;\;$(fb)} \\
\cline{3-9}
$500$GeV & (GeV) 
& Total & $|\gamma\gamma H|^2$& $|Z\gamma H|^2$ & $|$box$|^2$
& Int.$_{(\gamma\gamma H-Z\gamma H)}$  & Int.$_{(\gamma\gamma H-box)}$  
& Int.$_{(Z\gamma H-box)}$  
\\
\hline
$(P_e = 0 ;$ &
  80 & 1.85 & 1.49 & 0.203 & 0.0232 & 0.0765 & 0.0312 & 0.0205 \\
$ P_{\gamma} = 0)$ &
 100 & 1.99 & 1.61 & 0.216 & 0.0237 & 0.0820 & 0.0322 & 0.0212 \\
&120 & 2.22 & 1.81 & 0.238 & 0.0244 & 0.0912 & 0.0340 & 0.0224 \\
&140 & 2.68 & 2.20 & 0.280 & 0.0261 & 0.109  & 0.0377 & 0.0252 \\
\hline
$(P_e = -1 ;$ &
  80 & 2.72 & 1.49 & 0.244 & 0.0458 & 0.832  & 0.0544 & 0.0476 \\
$ P_{\gamma} = 0)$ &
 100 & 2.91 & 1.61 & 0.260 & 0.0466 & 0.892  & 0.0559 & 0.0494 \\
&120 & 3.24 & 1.81 & 0.286 & 0.0481 & 0.991  & 0.0586 & 0.0526 \\
&140 & 3.89 & 2.20 & 0.336 & 0.0513 & 1.18   & 0.0647 & 0.0590 \\
\hline
$(P_e = +1 ;$ &
  80 & 0.979& 1.49 & 0.162 &6.78E-04& -0.679 &8.02E-03&-6.68E-03 \\
$ P_{\gamma} = 0)$ &
 100 & 1.06 & 1.61 & 0.173 &7.06E-04& -0.728 &8.53E-03&-7.09E-03 \\
&120 & 1.19 & 1.81 & 0.190 &7.50E-04& -0.809 &9.31E-03&-7.70E-03 \\
&140 & 1.46 & 2.20 & 0.224 &8.18E-04& -0.965 & 0.0106 &-8.72E-03 \\
\hline
$(P_e = 0 ;$ &
  80 & 1.94 & 1.49 & 0.211 & 0.0344 & 0.179  &8.59E-03& 0.0108 \\
$ P_{\gamma} = -1)$ &
 100 & 2.10 & 1.61 & 0.226 & 0.0353 & 0.203  &9.64E-03& 0.0122 \\
&120 & 2.36 & 1.81 & 0.249 & 0.0369 & 0.240  & 0.0114 & 0.0145 \\
&140 & 2.87 & 2.20 & 0.294 & 0.0402 & 0.307  & 0.0151 & 0.0190 \\
\hline
$(P_e = 0 ;$ &
  80 & 1.76 & 1.49 & 0.195 & 0.0121 &-0.0255 & 0.0538 & 0.0302 \\
$ P_{\gamma} = +1)$ &
 100 & 1.88 & 1.61 & 0.207 & 0.0120 &-0.0385 & 0.0547 & 0.0301 \\
&120 & 2.08 & 1.81 & 0.227 & 0.0119 &-0.0578 & 0.0565 & 0.0303 \\
&140 & 2.48 & 2.20 & 0.266 & 0.0120 &-0.0896 & 0.0603 & 0.0313 \\
\hline
\end{tabular}

\end{center}
\caption{{\em Interference pattern between the $\gamma\gamma Z$,
$Z\gamma H$ and boxes contributions versus the 
$e$-beam and $\gamma$-beam polarizations, for 
$p^e_T > 10\,$GeV. 
}}
\label{tab53a}
\vspace*{-.3cm}
\end{table}
\begin{table}[t]
\vspace*{1cm}
\begin{center}
\hspace*{-1.5cm}
\begin{tabular}{|c|c||c|c|c|c|c|c|c|}
\hline
$\sqrt s =$& $m_H$
&\multicolumn{7}{c|}
{$\sigma(e\gamma\to e H, \;\; p^e_T > 100\,$GeV)$\;\;\;$(fb)} \\
\cline{3-9}
$500$GeV & (GeV) 
& Total & $|\gamma\gamma H|^2$& $|Z\gamma H|^2$ & $|$box$|^2$
& Int.$_{(\gamma\gamma H-Z\gamma H)}$  & Int.$_{(\gamma\gamma H-box)}$  
& Int.$_{(Z\gamma H-box)}$  
\\
\hline
$(P_e = 0 ;$ &
  80 & 0.516 & 0.265 & 0.149 & 0.0208 & 0.0393 & 0.0247 & 0.0177 \\
$ P_{\gamma} = 0)$  &
 100 & 0.542 & 0.278 & 0.158 & 0.0210 & 0.0415 & 0.0254 & 0.0182 \\
&120 & 0.584 & 0.301 & 0.171 & 0.0213 & 0.0449 & 0.0268 & 0.0192  \\  
&140 & 0.668 & 0.347 & 0.196 & 0.0221 & 0.0517 & 0.0298 & 0.0214  \\
\hline
$(P_e = -1 ;$ &
  80 & 0.996 & 0.265 & 0.179 & 0.0412 & 0.428  & 0.0424 & 0.0414 \\
$ P_{\gamma} = 0)$ &
 100 & 1.05  & 0.278 & 0.189 & 0.0414 & 0.451  & 0.0435 & 0.0428 \\
&120 & 1.13  & 0.301 & 0.205 & 0.0420 & 0.488  & 0.0457 & 0.453 \\
&140 & 1.29  & 0.347 & 0.235 & 0.0435 & 0.562  & 0.0505 & 0.506 \\
\hline
$(P_e = +1 ;$ &
  80 &0.0364 & 0.265 & 0.119 &5.13E-04& -0.349 &6.97E-03&-6.02E-03 \\
$ P_{\gamma} = 0)$ &
 100 &0.0381 & 0.278 & 0.126 &5.29E-04& -0.368 &7.37E-03&-6.37E-03 \\
&120 &0.0410 & 0.301 & 0.136 &5.54E-04& -0.399 &8.00E-03&-6.89E-03 \\
&140 &0.0471 & 0.347 & 0.157 &5.93E-04& -0.458 &9.05E-03&-7.75E-03 \\
\hline
$(P_e = 0 ;$ &
  80 & 0.592 & 0.265 & 0.157 & 0.0302 & 0.124  &7.47E-03& 8.53E-03  \\
$ P_{\gamma} = -1)$ &
 100 & 0.630 & 0.278 & 0.166 & 0.0306 & 0.137  &8.47E-03& 9.69E-03 \\
&120 & 0.692 & 0.301 & 0.180 & 0.0313 & 0.157  &0.0102  & 0.0116\\
&140 & 0.808 & 0.347 & 0.208 & 0.0328 & 0.192  &0.0136  & 0.0153 \\
\hline
$(P_e = 0 ;$ &
  80 & 0.441 & 0.265 & 0.142 & 0.0115 &-0.0456 & 0.0419 & 0.0269  \\
$ P_{\gamma} = +1)$ &
 100 & 0.454 & 0.278 & 0.149 & 0.0114 &-0.0542 & 0.0424 & 0.0268 \\
&120 & 0.477 & 0.301 & 0.161 & 0.0113 &-0.0671 & 0.0435 & 0.0268  \\
&140 & 0.527 & 0.347 & 0.184 & 0.0113 &-0.0883 & 0.0460 & 0.0275  \\
\hline
\end{tabular}

\end{center}
\caption{{\em Interference pattern between the $\gamma\gamma Z$,
$Z\gamma H$ and boxes contributions versus the 
$e$-beam and $\gamma$-beam polarizations, for 
$p^e_T > 100\,$GeV. 
}}
\label{tab53b}
\vspace*{-.3cm}
\end{table}

\begin{table}[b]
\begin{center}
\begin{tabular}{|c||c|c|c||c|c|c|}
\hline
$m_H=120$ GeV& \multicolumn{3}{|c||}{$ \sigma(e\gamma\to eH)$ fb}
             & \multicolumn{3}{|c|}{$ \sigma(e\gamma\to eb\bar b)$ fb} \\
\cline{2-7}
$\sqrt{s}=500$ GeV
&no $\theta_e$ cut&$\theta_e<90^\circ$&${\cal S}_{FB}$
&no $\theta_e$ cut&$\theta_e<90^\circ$&${\cal S}_{FB}$ \\
\hline
$P_e=0$ & 0.404 & 0.362 & 0.320 & 0.634 & 0.281 & -0.072 \\
$P_e=-1$& 0.780  & 0.699 & 0.618 & 0.961 & 0.433 & -0.095 \\
$P_e=1$ & 0.0263& 0.0258& 0.0253& 0.304 & 0.126 & -0.052 \\
\hline
\end{tabular}
\end{center}
\caption{{\em Forward-backward asymmetry in the
electron scattering angle,
${\protect\theta_e}$,  for different $e$-beam polarizations. 
The total cross section, the forward cross section and 
the difference of the forward and backward cross sections 
${\protect[{\cal S}_{FB}=\sigma(\theta_e<90^\circ)-\sigma(\theta_e>90^\circ)]}$
for the signal and the irreducible background, are presented.
The kinematical cuts ${\protect\theta_{b-beam}>18^\circ}$, $P_t^e>100$ GeV and 
${\protect\mh-\delbb< \mbb <\mh+\delbb}$, with 
${\protect\delbb=3}$ GeV, are applied.
}}
\label{tab71}
\end{table}



\end{document}